\documentclass[12pt]{article}
 
\usepackage{epsfig}
\usepackage{psfig}

\begin{document}
\begin{center}
{\Large\bf
Topologically Unquenched QCD: Prospects\\
from$\;$an$\;$explorative$\;$study$\;$in$\;$2-flavour-QED(2)}
\end{center}
\vspace{2pt}

\begin{center}
{\bf Stephan D\"urr}
\vspace{2pt}\\
{\sl Paul Scherrer Institut, Theory Group}\\
{\sl 5232 Villigen PSI, Switzerland}
\vspace{1pt}\\
{\tt stephan.duerr@psi.ch}
\end{center}
\vspace{-2pt}

\begin{abstract}
The concept of simulating a variant of QCD with sea quarks which interact
with the gluon configuration only via global gluonic quantities like
$\int\!F\tilde F\,dx$ and $\int\!F F\,dx$ is tested for the case of the massive
2-flavour Schwinger model. It is found to amount to an importance sampling
method which generates --at essentially the costs of a quenched run-- an
ensemble in between a full and a quenched one, thus providing a competitive
alternative to the quenched approximation.
\end{abstract}
\vspace{2pt}


\section{Introduction}

Simulations of lattice QCD use either the quenched or the partially
quen\-ched approximations, where internal fermion loops are omitted or at least
suppressed by giving the dynamical (`sea') quarks a mass heavier than that of
the propagating (`current') quarks \cite{Hamber, Eicker}.
Such an approximation turns out to be necessary, since presently available
update algorithms with dynamical fermions tend to slow down rather dramatically
if the fermion mass is taken light.
From a physical point of view this unwanted feature does not come as a
surprise, as it reflects the very nonlocal nature of the fermion determinant.
The quenched and the partially quenched approximations amount thus to a
suppression or reduction of the nonlocal self-interaction of the gluon
configuration brought by the sea quarks.
Though this seems reasonable, there are specific problems brought by the
quenched and --to a milder extent-- the partially quenched approximations:
It has been noted very early that quenched QCD lacks reflection positivity
\cite{Morel} which is just the Euclidean analog of unitarity.
This means that Green's functions computed in the quenched theory cannot be
rotated back to Minkowski space and hence there is no Hamiltonian formulation.
Attempts to construct a low-energy theory adapted to the quenched and partially
quenched approximations have shown that these constructs give rise to
unphysical degrees of freedom which persist in low energy observables and
render the whole theory sick, if the chiral limit is taken \cite{QuenchedQCD,
PartiallyQuenchedQCD}.
The computational overhead, though, of full QCD with
phenomenological quark masses is so dramatic that --without tremendous
progress at the algorithmic front-- both the quenched and the partially
quenched approximations are likely to stay with us for the foreseeable future.

\section{Outline of the method}

Since the ultimate goal --simulating full QCD-- is fixed, any potential
alternative to the standard path `quenched -- partially quenched -- full QCD'
can only imply a different starting point and/or a different way to
interpolate between the measure one starts with and the full one.
Ideally, one would like to begin with a measure which keeps costs per
configuration essentially at the quenched level, but tries to incorporate
as much of the fermion dynamics brought by the functional determinant as
possible. A specific attempt in this direction goes with the name
`topologically unquenched QCD' \cite{TUQCD}. The basic idea is to split the
functional determinant into two factors
\begin{equation}
\det(D\!\!\!\!/+\!m)=
\det(D\!\!\!\!/+\!m)_{\rm guess}\cdot
{\det(D\!\!\!\!/+\!m)\over\det(D\!\!\!\!/+\!m)_{\rm guess}}
\label{detfac}
\end{equation}
the first of which is computationally cheap and as good an approximation to the
whole determinant as possible. This factor is then absorbed into the measure.
On the other hand, the right hand factor is too expensive to be determined for
each link update and henceforth this second factor is either discarded of or
just included into the observable\footnote{If this is possible at all -- for a
discussion see below.}.
Of course, this concept can only work if it is possible to come up with a
simple (i.e.\ computationally cheap to evaluate) and reasonably accurate choice
for the first factor.

In the original publication \cite{TUQCD} it was proposed to define the first
factor as the determinant of a suitably chosen background which shares with
the actual configuration nothing but the overall topological charge.
Below two modifications are proposed, but the basic concept remains the same:
The first factor in (\ref{detfac}) is nothing but a statistical guess of the
total determinant which uses as input only global gluonic quantities.
The first suggested change concerns the size of the database the guess is
based on. Originally the idea was to compute the first factor in each
topological sector once -- on a well chosen configuration \cite{TUQCD}.
In \cite{LAT99} it was already mentioned that in practice it proves useful to
construct the guess as the geometric average of the determinants of few
configurations which are quite typical both in their gauge action and in
the logarithm of their determinant.
Here it is suggested to feed the database with the combination
$\int\!F\tilde F\,dx,-\!\log(\det(D\!\!\!\!/+\!m))$ for each `complete'
configuration, i.e.\ after the entire lattice has been swept through a few
times. The point is that the `complete' configurations are those which are used
for measurements, i.e.\ those in which the determinant emerges as a byproduct
of the Dirac operator inversion and thus comes for free\footnote{Except for the
thermalization phase, which now also requires inversions of $D\!\!\!\!/+\!m$.}.
The second suggested change concerns the argument being used to `read out' the
guess for the first factor in (\ref{detfac}) from the database.
The original proposal was to make a guess based on $\int\!F\tilde F\,dx$ only.
In the present work it is eventually proposed to use the combined input from
$\int\!F\tilde F\,dx$ and $\int\!F F\,dx$.
The basic idea is still, when a specific link update is proposed, {\em not to
compute\/} the change in the determinant implied, but rather to use the
databases knowledge regarding the determinant ratio the proposed change in the
gluon configuration brings {\em on average\/}.
The original task to compute, for each local link update, a {\em global
fermionic\/} quantity is thus traded for the task to compute the actual change
in one or two {\em global gluonic\/} quantities.
The latter, however, reduces to the stunningly simple task to determine, for
each link update, one or two {\em local gluonic\/} quantities, since
$\int\!F\tilde F\,dx$ and $\int\!F F\,dx$ can easily be kept track of by
computing, for each proposed update, the corresponding contribution from the
staple\footnote{Here an unimproved action is assumed. In the case of a gluon
action from `fuzzed' links or with next-to-nearest neighbor couplings the
correspondingly `fuzzed' staple --an object constructed by stapling each
link in a conventional staple-- must be looked at.}
around the link under consideration, before and after it is modified.
In other words, the connection between the desired global quantity
and the readily accessible local gluonic quantity is made via a two-step
procedure: The database stores the correlation between
$-\!\log(\det(D\!\!\!\!/+\!m))$ and $\int\!F\tilde F\,dx, \int\!F F\,dx$, i.e.
on a global-to-global basis. In an updating sweep the actual value of these
integrated quantities is constantly kept track of~-- a task which, at any time,
requires only looking at the lattice versions of $F\tilde F(x)$ and $FF(x)$ in
the immediate neighborhood of the link under consideration.
The price to pay is that the guess for the change in the determinant the
proposed link update would bring about is not accurate but only correct on
average.

From this outline it is clear that the concept can only work if there is
a correlation --in fact, for practical purposes, a sufficiently strong
correlation-- between the logarithm of the determinant and, on
the other side, the combination of topological index and gluonic action of a
configuration generated in a Monte Carlo process.
In addition, a few practical issues must be resolved, e.g. the question how
each configuration can be assigned a topological index in a such a way that
the costs (in terms of CPU time) for its determination are small compared to
the savings from the fewer determinant evaluations.

\section{Outline of the testbed}

While the idea of `topologically unquenched QCD' was originally put
forth~\cite{TUQCD} and successively advertised~\cite{TUQCDreview} with a strong
focus on theoretical aspects (merely arguing that in the continuum a
correlation between the functional determinant and the topological index is to
be expected from the Instanton Liquid Model~\cite{SchaferShuryak}), a fair
assessment of its potential can only come from a practical test.
Since it is desirable to have direct comparison to simulation data obtained
with an exact algorithm in the full theory, a first account may well stem from
a test implementation in a suitable model field theory.

A toy theory where the issue of `topological unquenching' can be studied
without having to worry about many technical aspects of the implementation
(which do indeed matter in QCD) is the massive multi-flavour Schwinger model,
i.e.\ QED(2) with $N_{\! f}\!\geq\!2$ light degenerate fermions.
The physics of this theory resembles in many aspects QCD(4) slightly {\em above
the phase transition\/}  (see Refs.~\cite{SchwingerModel} and references
cited in the second of them), i.e.\ the chiral condensate vanishes (in the
chiral limit) and the expectation value for the Polyakov loop is real and
positive.
Like in QCD the gauge fields in the continuum version of the Schwinger model
quantized on the torus fall into topologically distinct classes
\cite{SachsWipf}, i.e.\ they can be attributed a topological index.
The latter agrees with the number of left- minus right-handed zero-modes of the
massless Dirac operator, i.e.\ the index theorem holds true.
A closer look reveals, however, that there is no precise analog of the
instanton, i.e.\ no topologically nontrivial solution which is both a local
minimum of the classical action and localized in space-time. There are
`instantons', i.e.\ topologically nontrivial minima of the classical action,
but they are completely delocalized. On the other hand, there are `vortices',
i.e.\ topologically nontrivial localized objects, but they do not represent
local minima of the classical action.
An obvious consequence for the intended test is that `topological unquenching'
can only work here, if it relies only on very general features, shared by a
broad class of QCD inspired theories.
\enlargethispage{12pt}

In order to test the `topological unquenching' idea, I have chosen to
implement the Schwinger model with 2 dynamical staggered fermions both with
the full and with the `topologically unquenched' measure, as well as the
quenched approximation. In all three cases the unimproved gauge action
$S_{\rm gauge}\!=\!\beta\sum(1\!-\cos(\theta_{\rm plaqu}\!))$ has been used.
To compare the three theories to each other 2000/4000/4000 (full/ topologically
unquenched/ quenched) independent configurations have been generated on a
lattice with volume $V\!=\!16\times 8$ with periodic/thermal boundary
conditions.
The common coupling is $\beta\!=\!1/g^2\!=\!4.0$ and the fermion mass is
$m\!=\!0.05$ (everything in lattice units).
These values are chosen such that the `pion' (pseudo-scalar iso-triplet) has a
mass\footnote{Note that $g$ has the dimension of a mass and, unlike in QCD,
the power by which $M_\pi$ depends on $m$ is specific for $N_{\! f}\!\!=\!2$,
the general formula is $M_\pi\propto m^{N_{\! f}/(N_{\! f}+1)}$.}
$M_\pi\!\simeq\!2.066\;g^{1/3}\;m^{2/3}\!\simeq\!0.2226$ as to fit into the
box. The formula used is the prediction by the bosonized (strong-coupling)
version of the theory (see Refs.\ cited in \cite{SchwingerModel}), which seems
adequate, as $m/g\!\simeq\!0.1\!\ll\!1$.
Since the staggered Dirac operator is represented by a small ($128\times 128$)
matrix with 5 entries per row or column, I have decided to compute the
determinants exactly, using the routines ZGEFA and ZGEDI from the LINPACK
package. Finally, since the staggered formulation has a remnant degeneracy, a
square root (in two dimensions) must be taken to get an object associated with
$\det(D\!\!\!\!/+\!m)$ in the continuum.
For the topological charge, I have implemented both the L\"uscher geometric
definition $\nu_{\rm geo}\!=\!{1\over2\pi}\sum\log(U_{\rm plaqu})$ and the
rescaled naive $\nu_{\rm nai}\!=\!\kappa_{\rm nai}\sum\sin(\theta_{\rm plaqu})$
with $\kappa_{\rm nai}\!=\!1/(1-\langle S_{\rm gauge}\rangle/\beta
V)$~\cite{SmiVin}.
A configuration is assigned an index and used for measurements only if
the geometric and the naive definition, after rounding to the nearest integer,
agree -- which, at $\beta\!=\!4.0$, holds true for more than $99\%$ of all
configurations. This fraction being so high means that in practice an
assignment can be done {\em without cooling\/} by simply taking $\nu_{\rm nai}$
and rounding it to the nearest integer $\nu$. Of course, this means that some
configurations are assigned an index which, strictly speaking, aren't
continuum-like enough as to have one, but, since the latter fraction is so
tiny, such a simple definition may be tried first\footnote{Implications from
inaccurate or arbitrary index assignments will be discussed below.}.

\section{Correlations in the full theory}

\begin{figure}
\epsfig{file=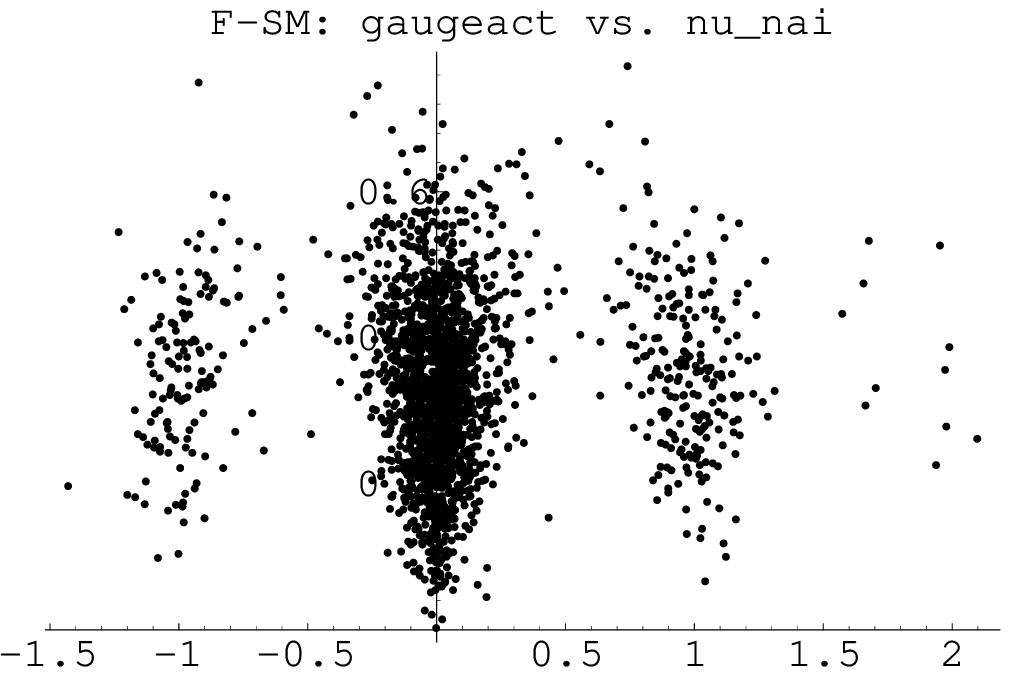,height=3.5cm,width=6.0cm,angle=0}
\hfill
\epsfig{file=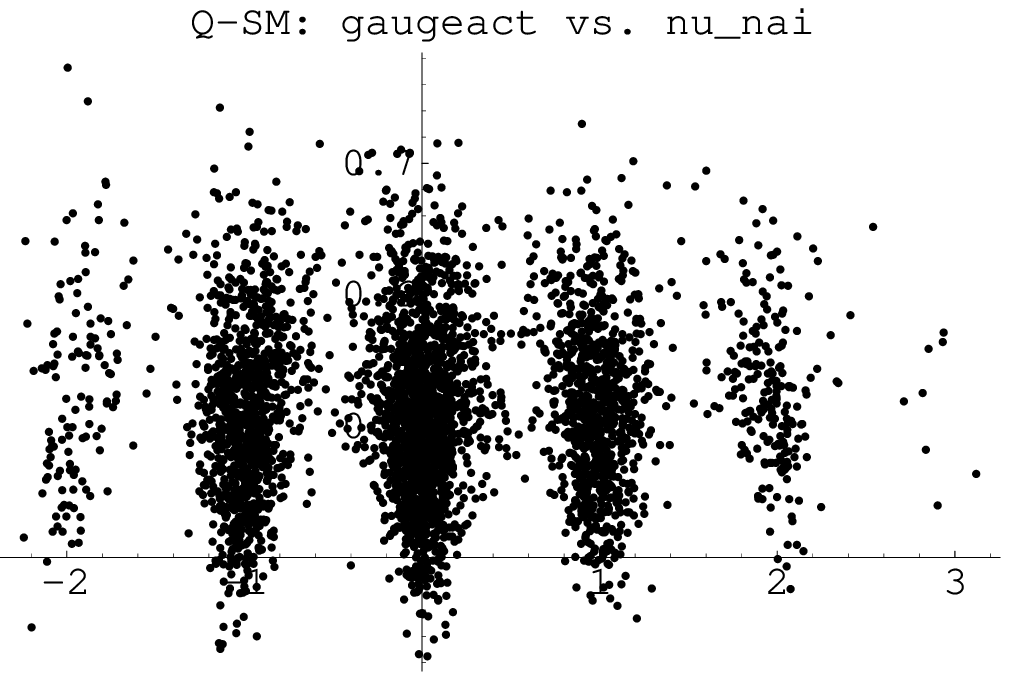,height=3.5cm,width=6.0cm,angle=0}
\vspace{3mm}
\\
\epsfig{file=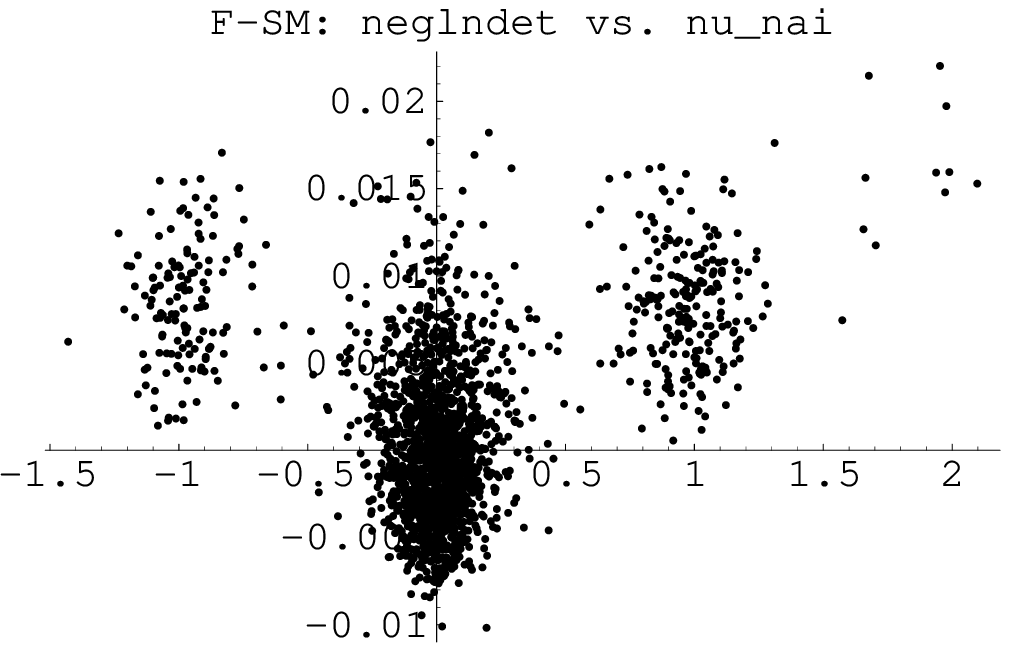,height=3.5cm,width=6.0cm,angle=0}
\hfill
\epsfig{file=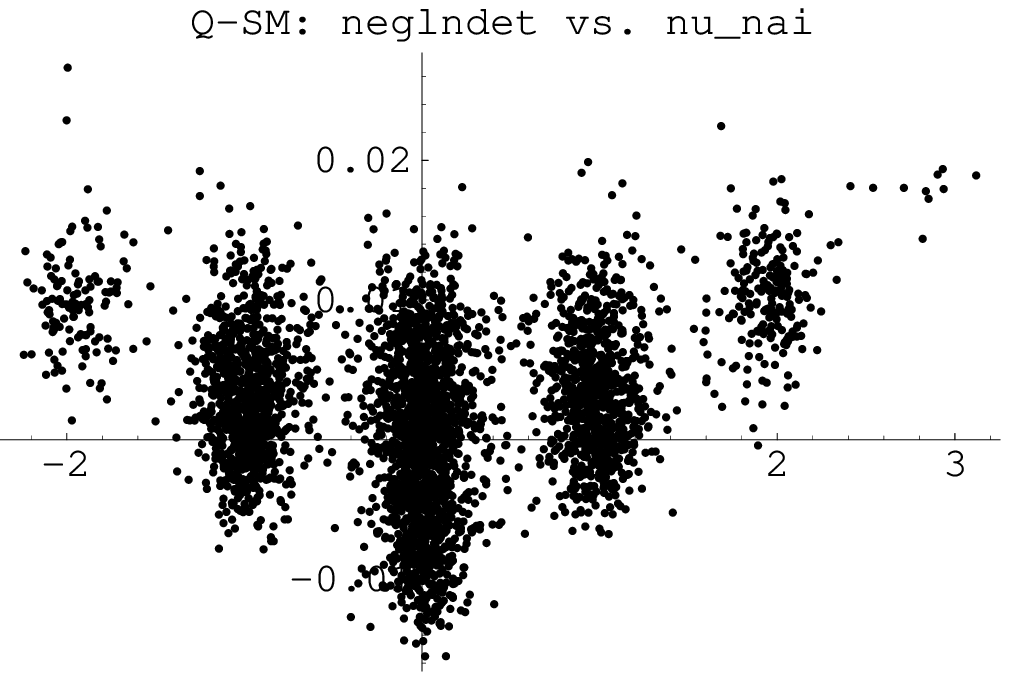,height=3.5cm,width=6.0cm,angle=0}
\vspace{3mm}
\\
\epsfig{file=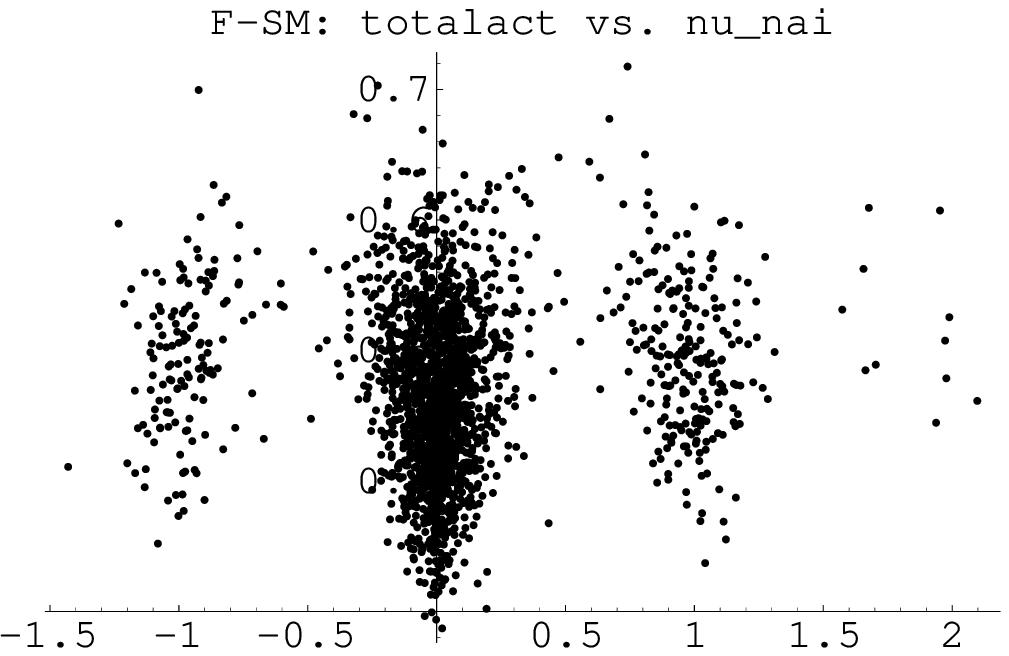,height=3.5cm,width=6.0cm,angle=0}
\hfill
\epsfig{file=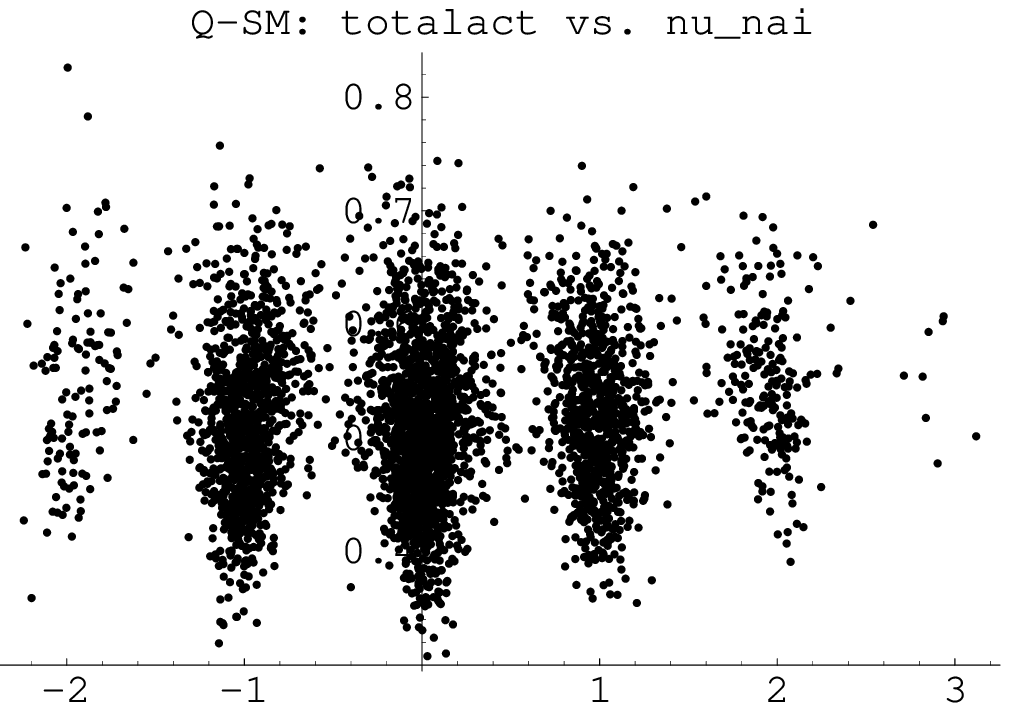,height=3.5cm,width=6.0cm,angle=0}
\vspace{-3mm}
\caption{\sl\small
Scatter plots of $S_{\rm gauge}$, $-\!\log(\det(D\!\!\!\!/+\!m))$ per
continuum flavour and $S_{\rm tot}$ [for two flavours] as a function of
$\nu_{\rm nai}$ in the 2-flavour-theory with $\beta\!=\!4.0, m\!=\!0.05$
(left) and in the quenched theory (right).}
\end{figure}
\begin{figure}
\epsfig{file=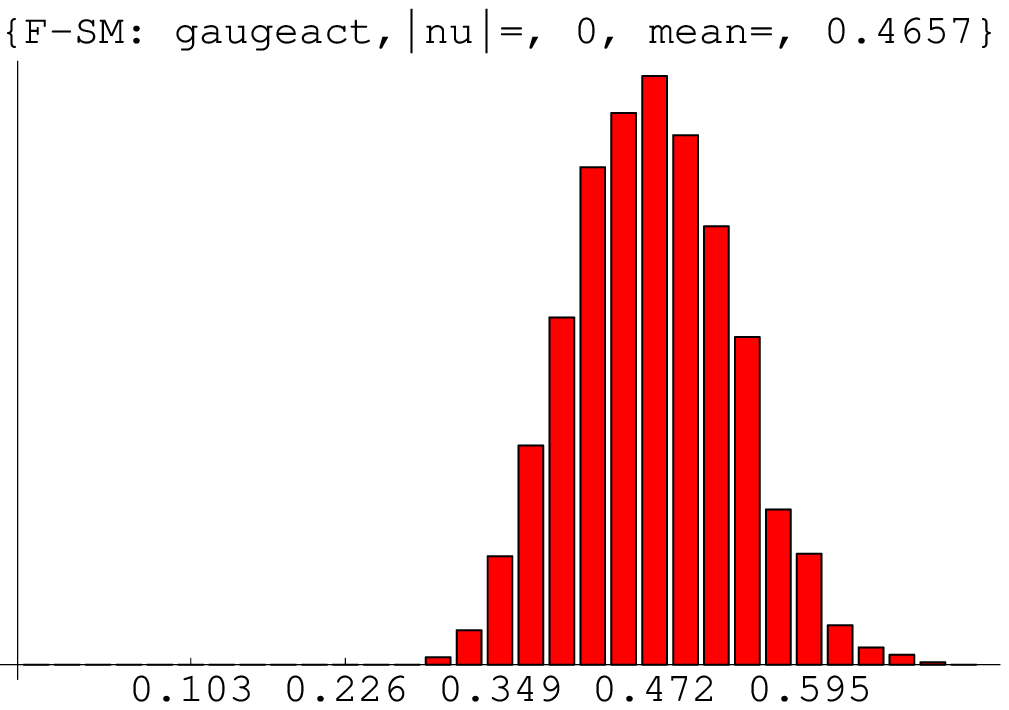,height=3.0cm,width=4.4cm,angle=0}
\hspace{0.5mm}
\epsfig{file=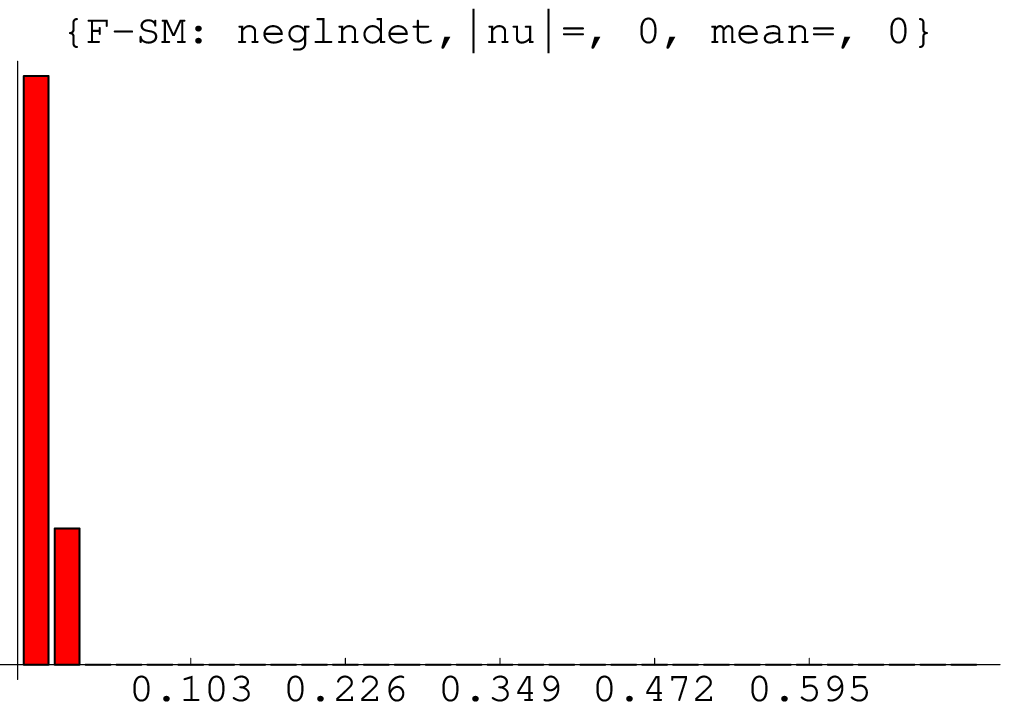,height=3.0cm,width=4.4cm,angle=0}
\hspace{0.5mm}
\epsfig{file=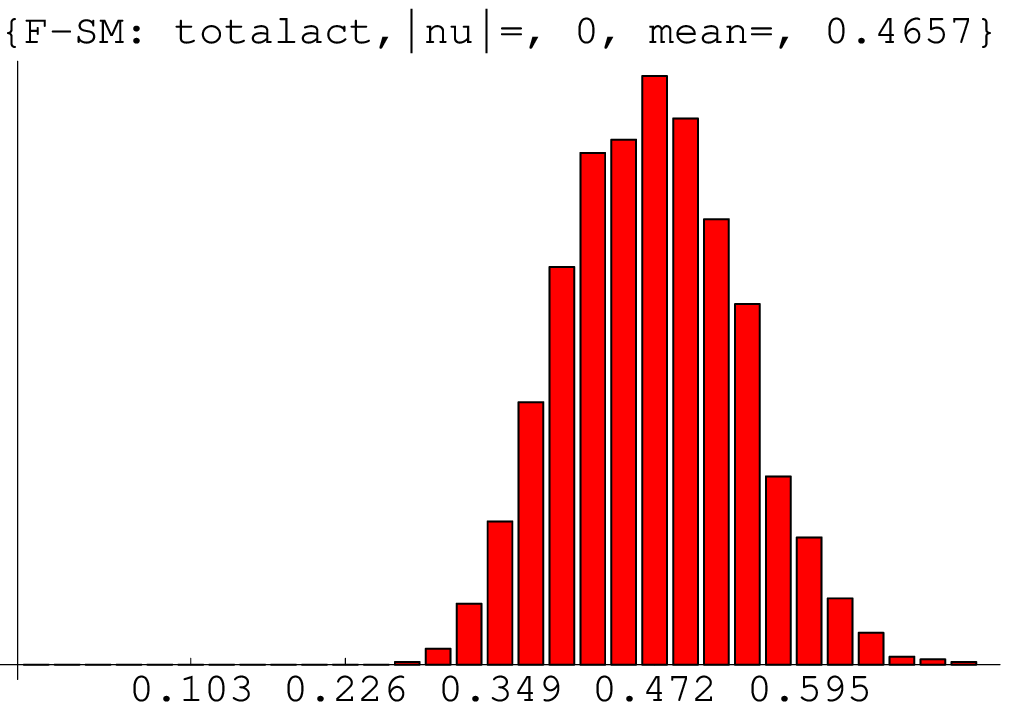,height=3.0cm,width=4.4cm,angle=0}
\vspace{-3mm}
\\
\epsfig{file=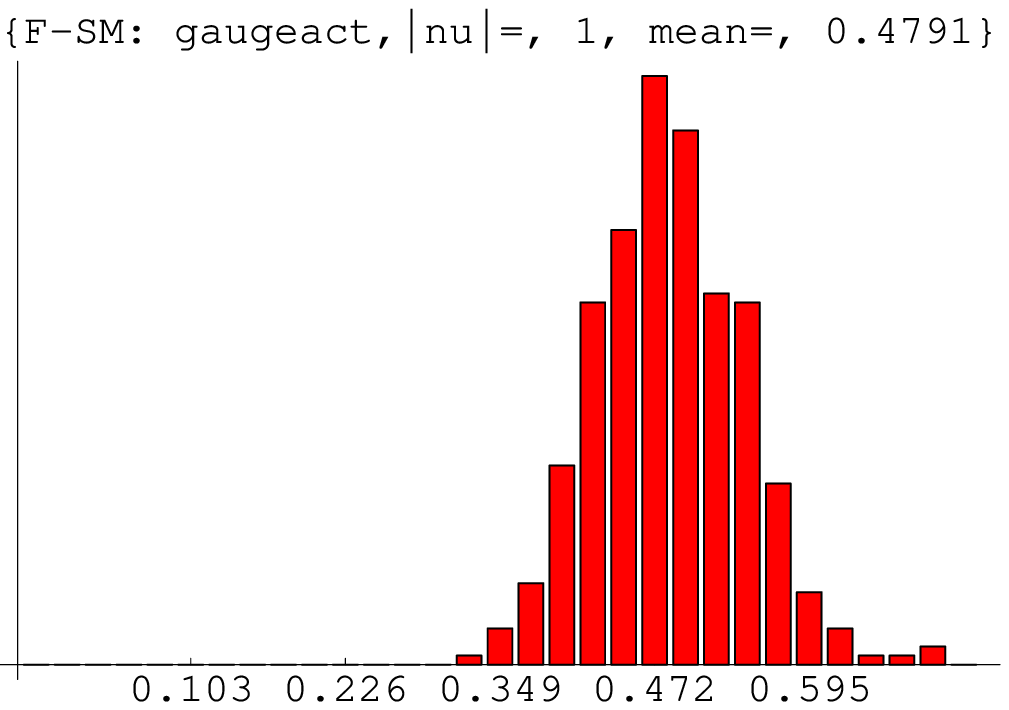,height=3.0cm,width=4.4cm,angle=0}
\hspace{0.5mm}
\epsfig{file=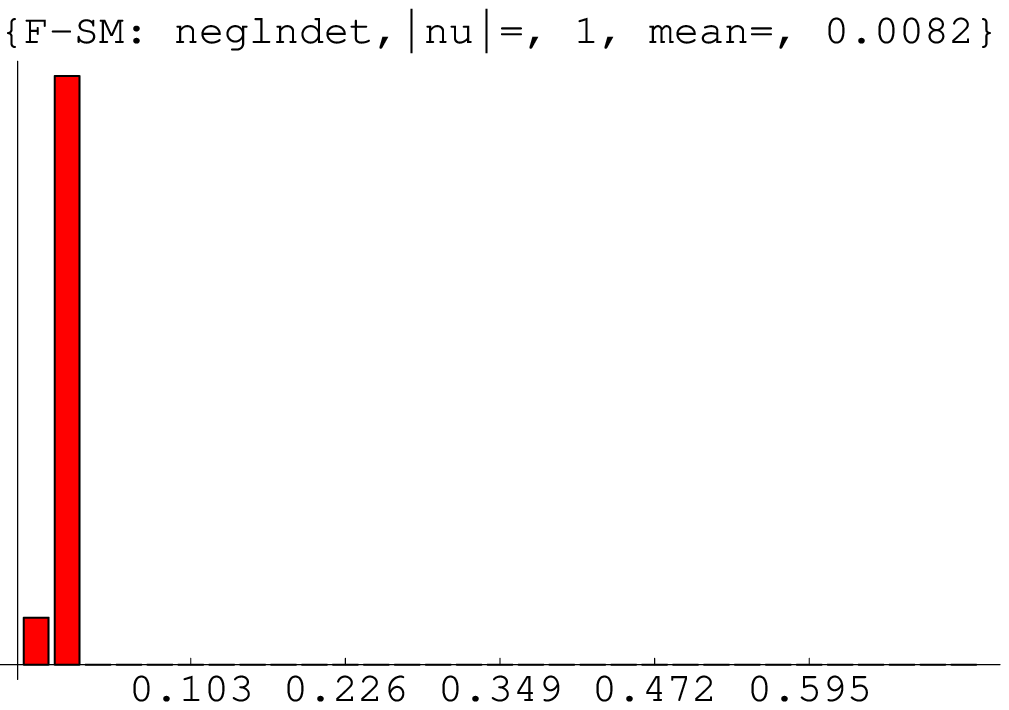,height=3.0cm,width=4.4cm,angle=0}
\hspace{0.5mm}
\epsfig{file=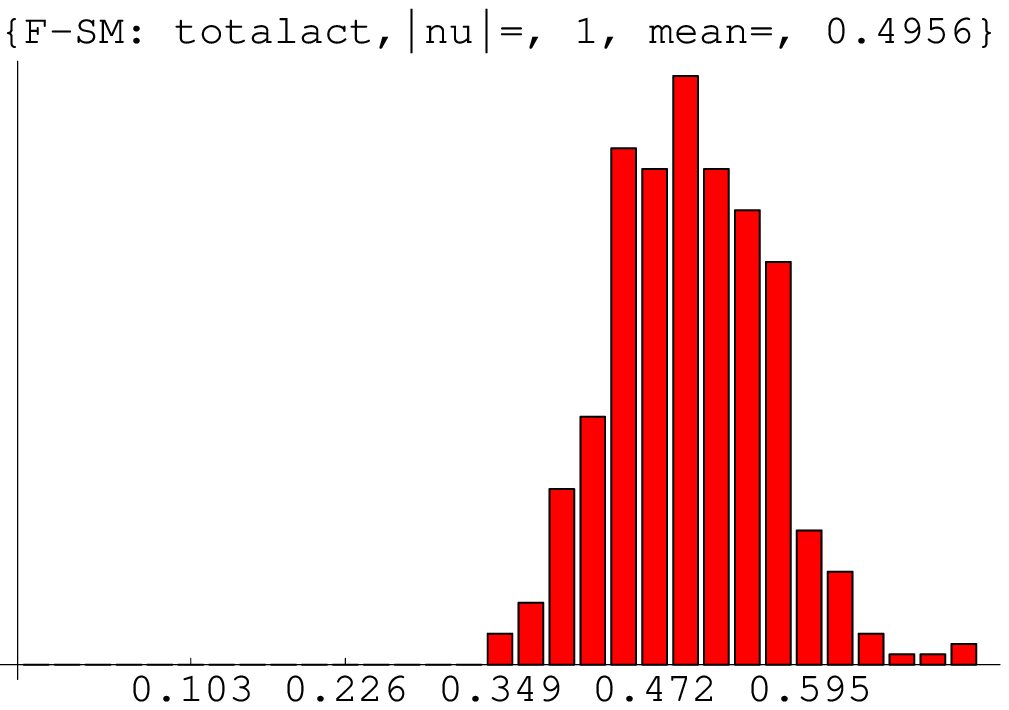,height=3.0cm,width=4.4cm,angle=0}
\vspace{-3mm}
\\
\epsfig{file=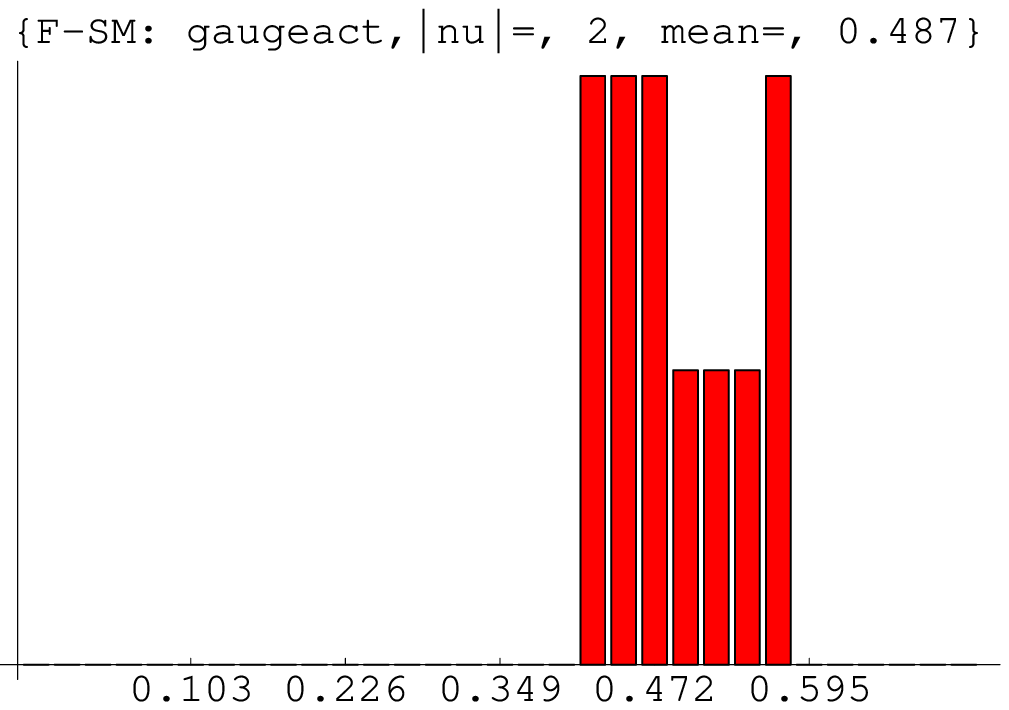,height=3.0cm,width=4.4cm,angle=0}
\hspace{0.5mm}
\epsfig{file=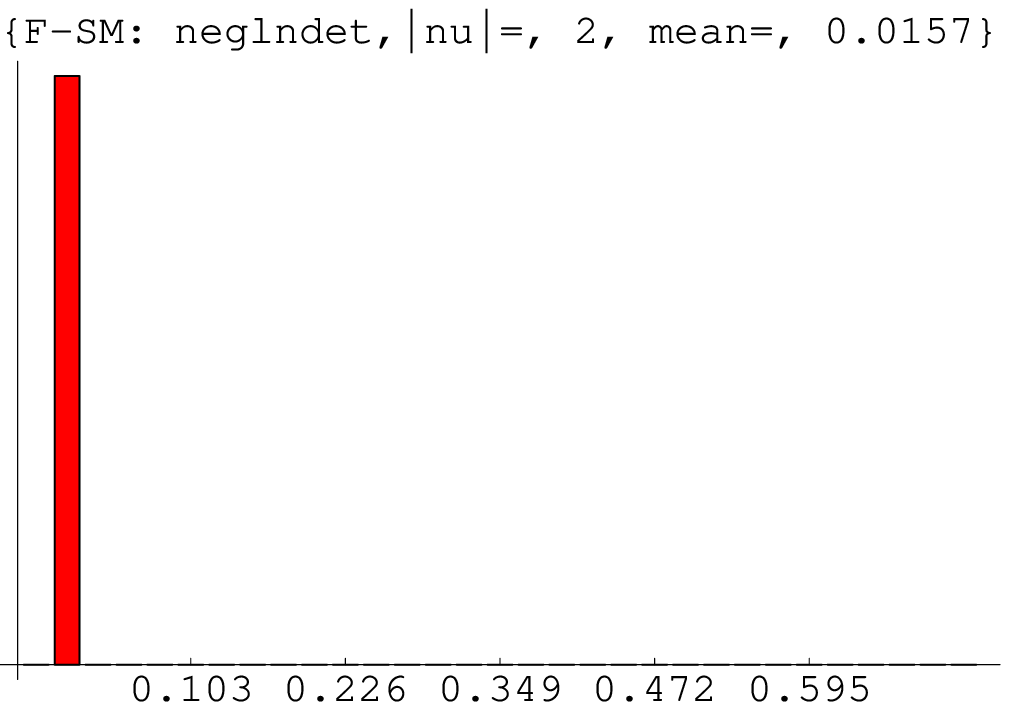,height=3.0cm,width=4.4cm,angle=0}
\hspace{0.5mm}
\epsfig{file=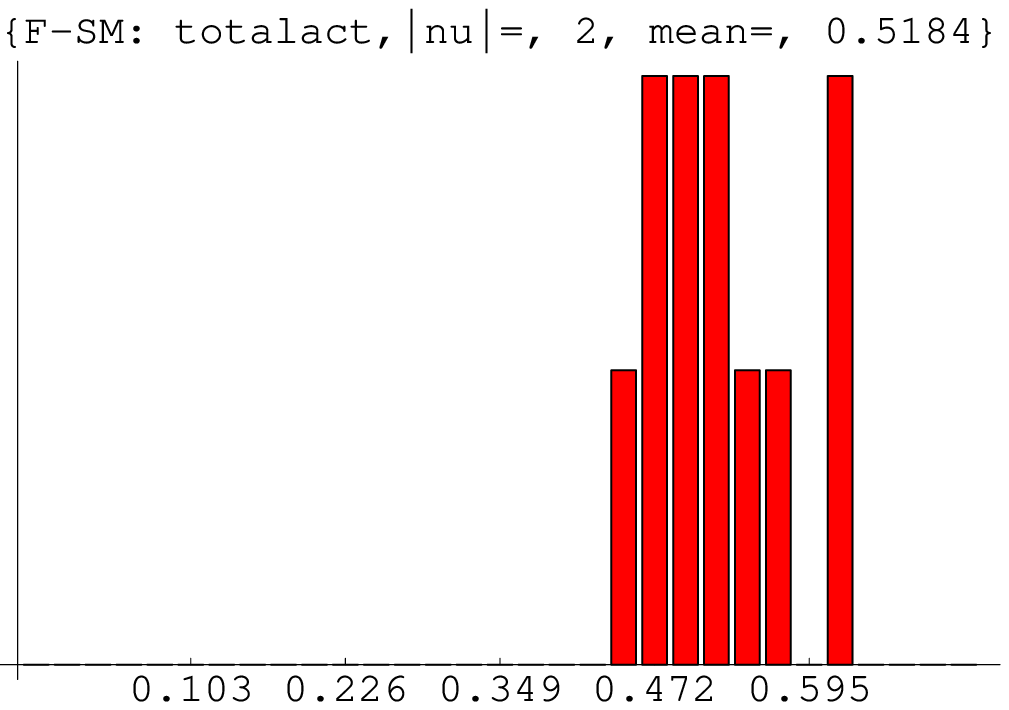,height=3.0cm,width=4.4cm,angle=0}
\vspace{-6mm}
\caption{\sl\small
Distribution of $S_{\rm gauge}$, $-\!\log(\det(D\!\!\!\!/+\!m))$ per continuum
flavour and $S_{\rm tot}$ for $|\nu|\!=\!0,1,2$ in the full 2-flavour-theory
($\beta\!=\!4.0, m\!=\!0.05$). For details see text.}
\end{figure}
\begin{figure}
\vspace{3mm}
\epsfig{file=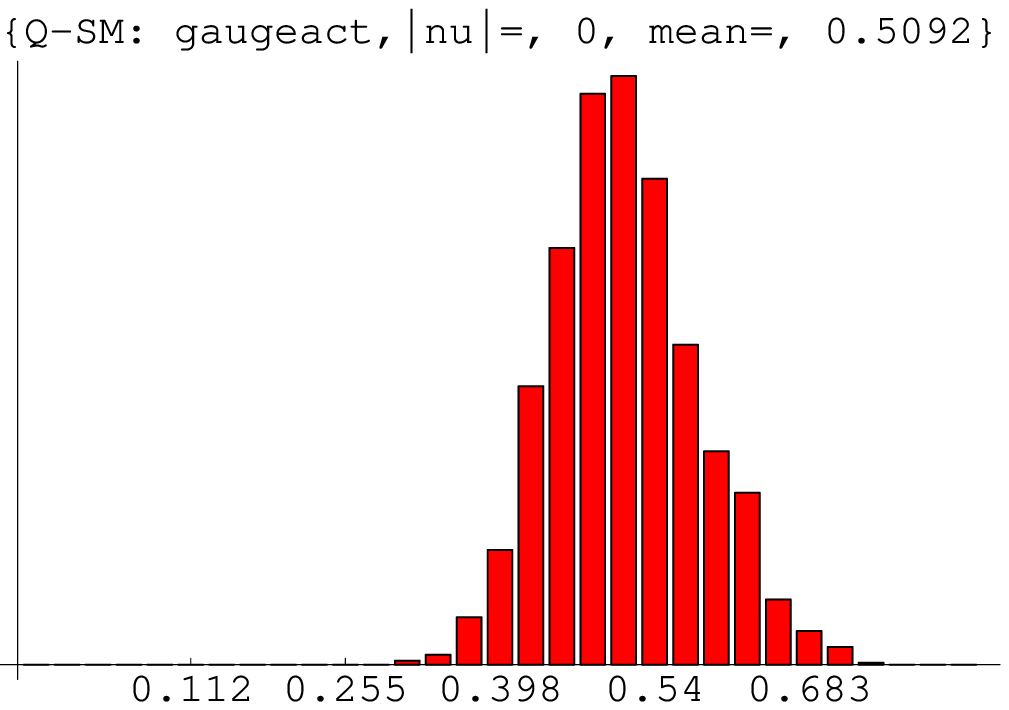,height=3.0cm,width=4.4cm,angle=0}
\hspace{0.5mm}
\epsfig{file=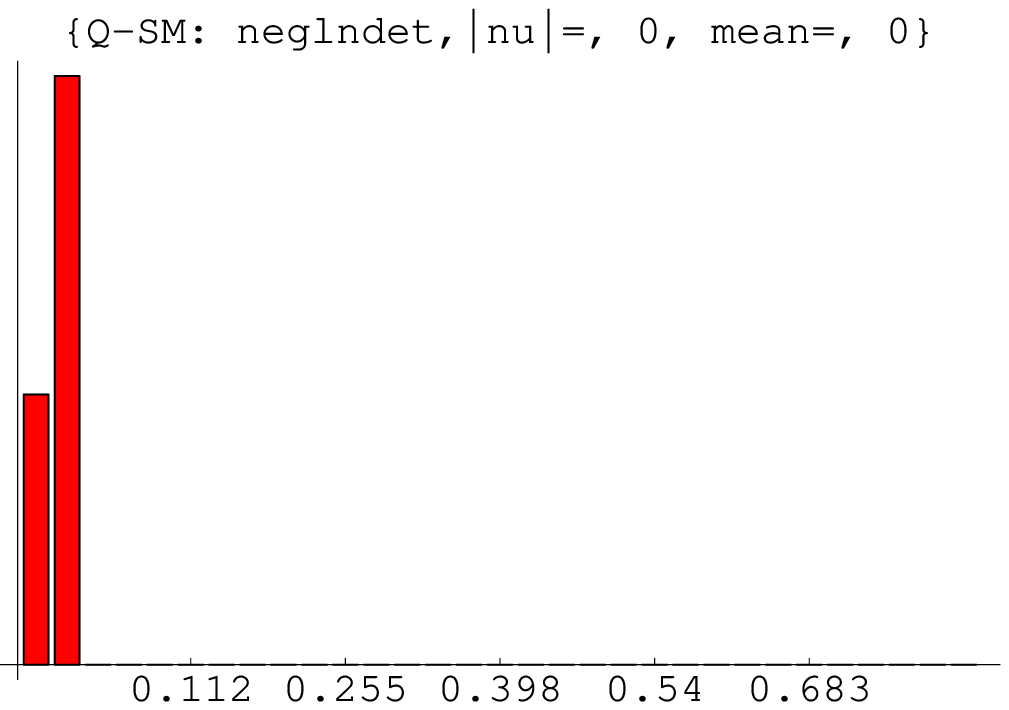,height=3.0cm,width=4.4cm,angle=0}
\hspace{0.5mm}
\epsfig{file=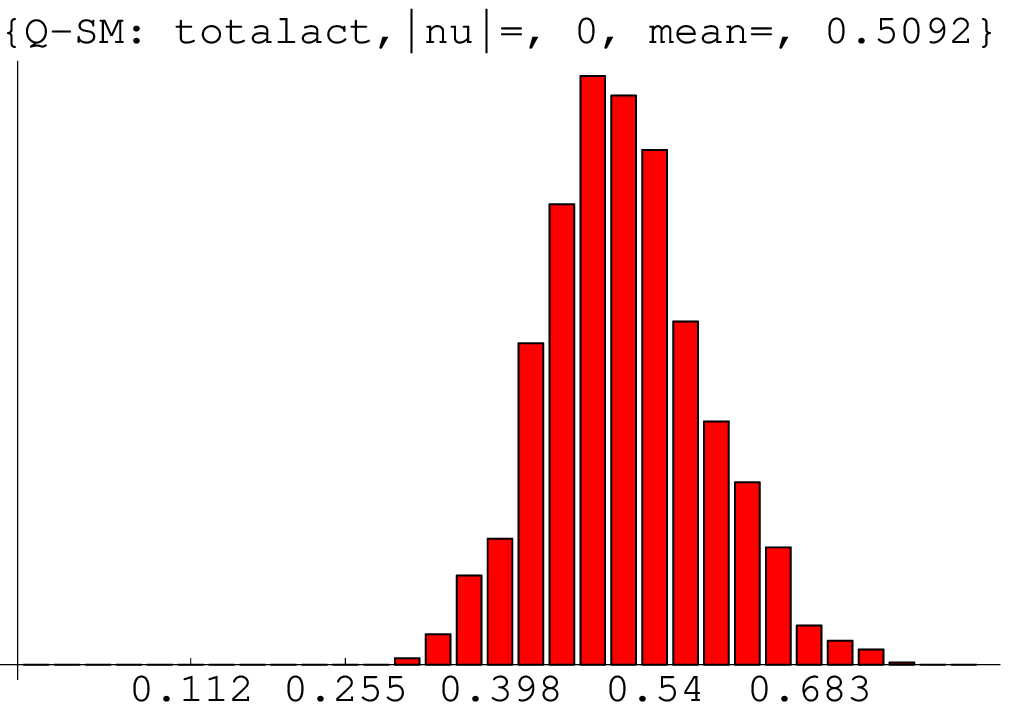,height=3.0cm,width=4.4cm,angle=0}
\vspace*{-3mm}
\\
\epsfig{file=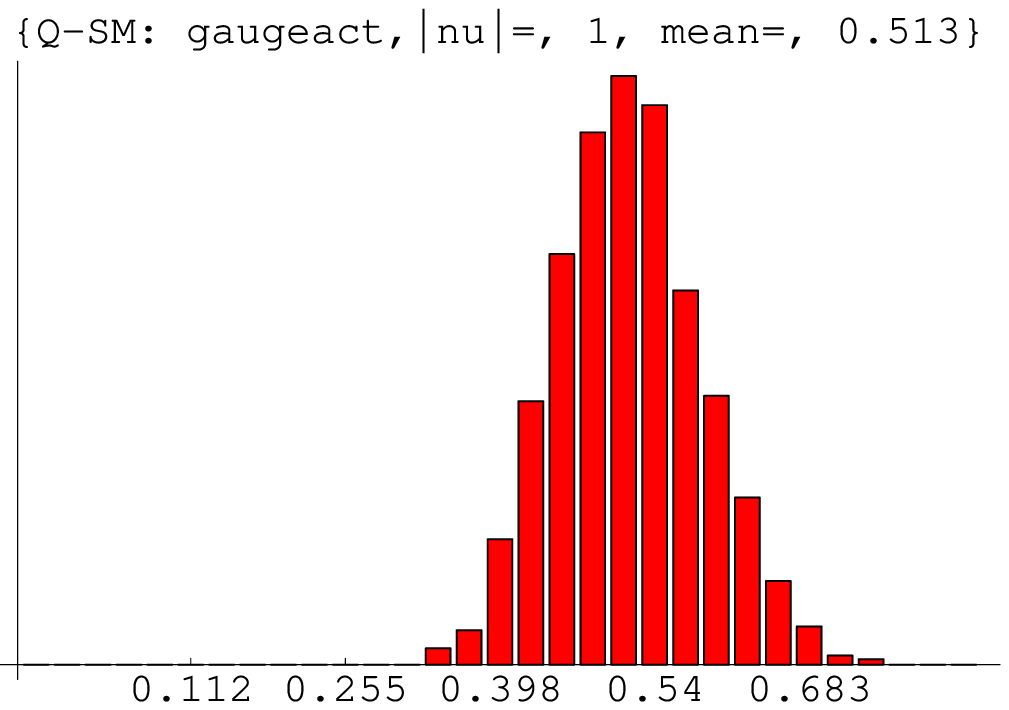,height=3.0cm,width=4.4cm,angle=0}
\hspace{0.5mm}
\epsfig{file=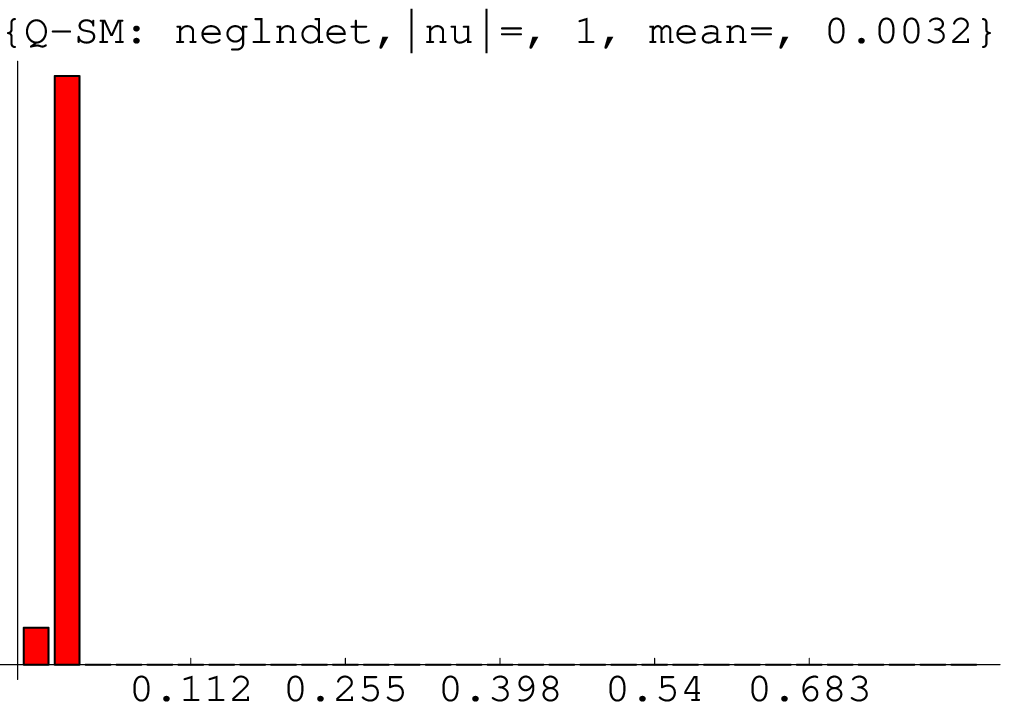,height=3.0cm,width=4.4cm,angle=0}
\hspace{0.5mm}
\epsfig{file=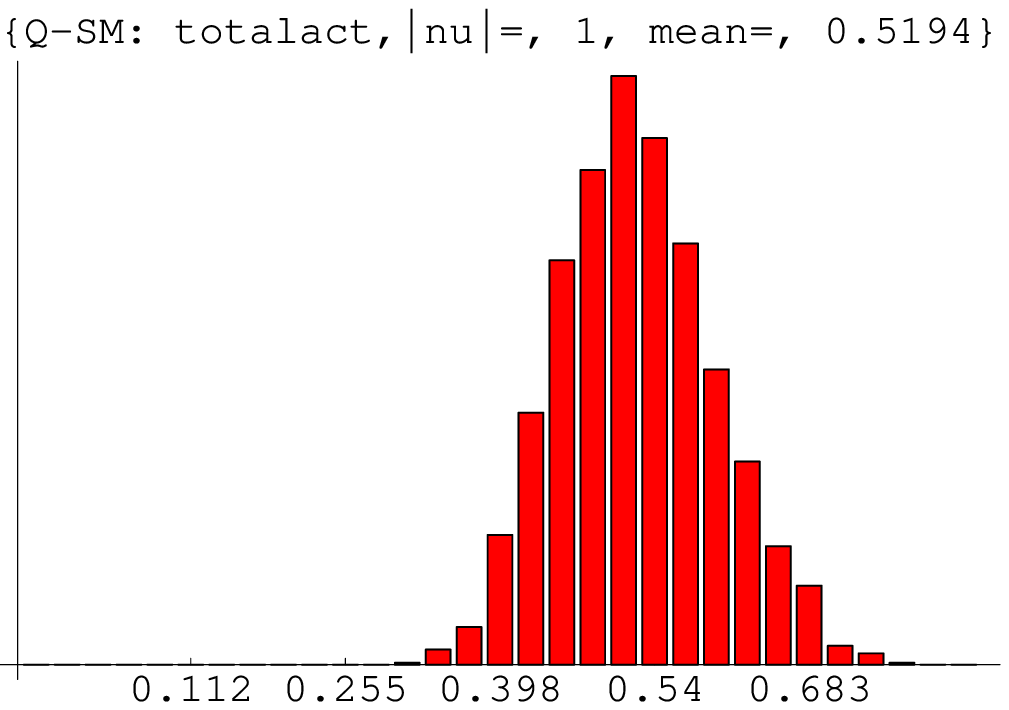,height=3.0cm,width=4.4cm,angle=0}
\vspace{-3mm}
\\
\epsfig{file=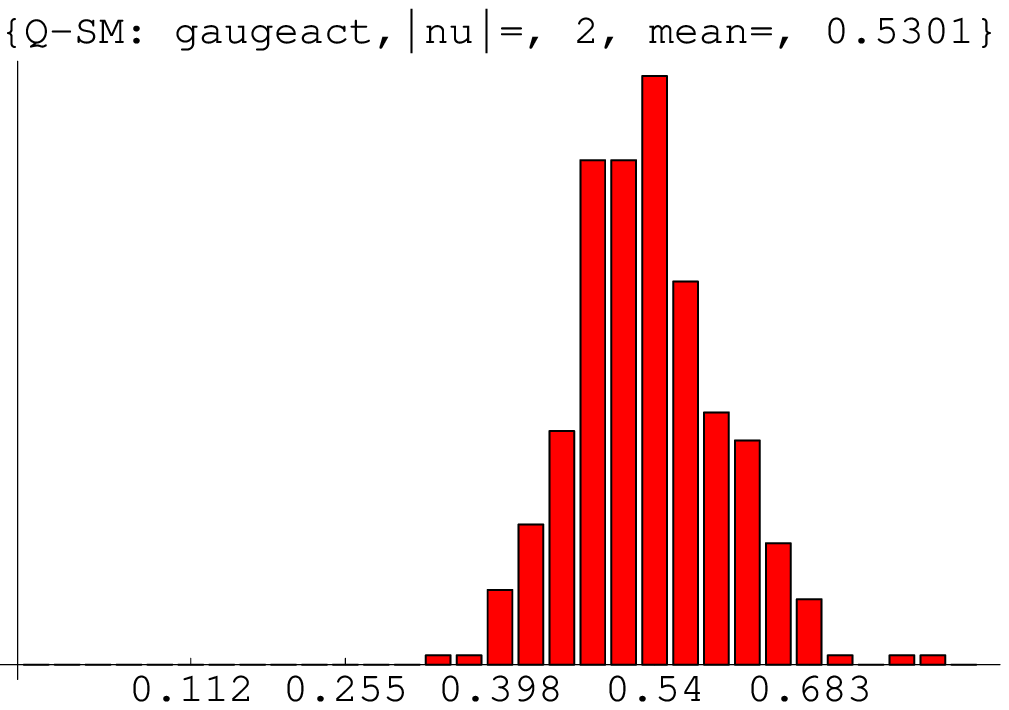,height=3.0cm,width=4.4cm,angle=0}
\hspace{0.5mm}
\epsfig{file=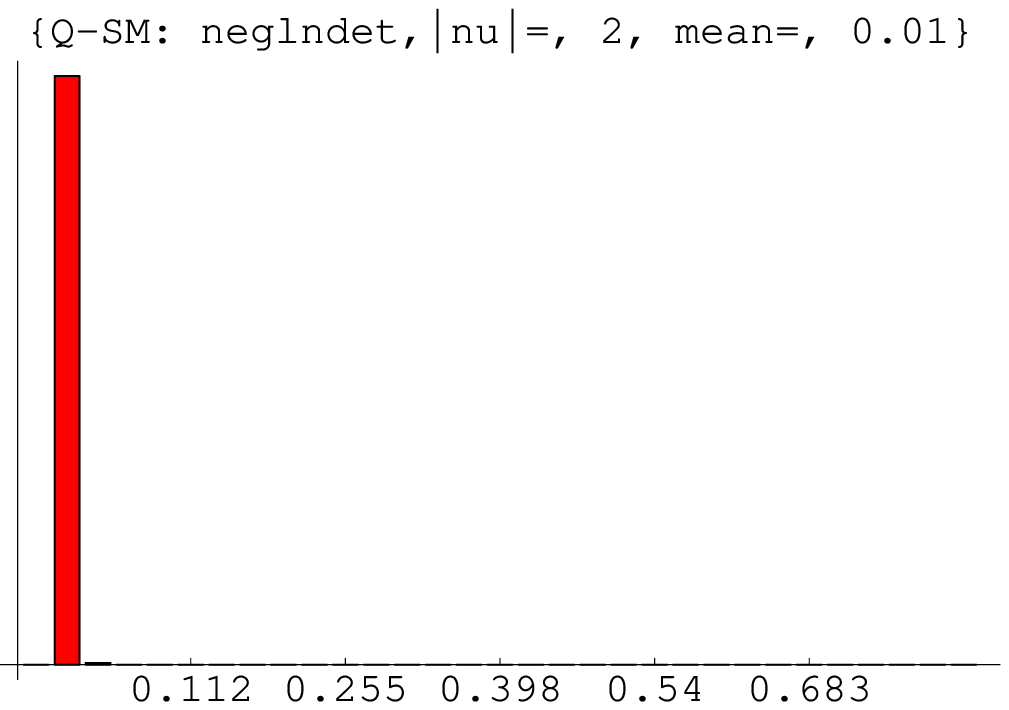,height=3.0cm,width=4.4cm,angle=0}
\hspace{0.5mm}
\epsfig{file=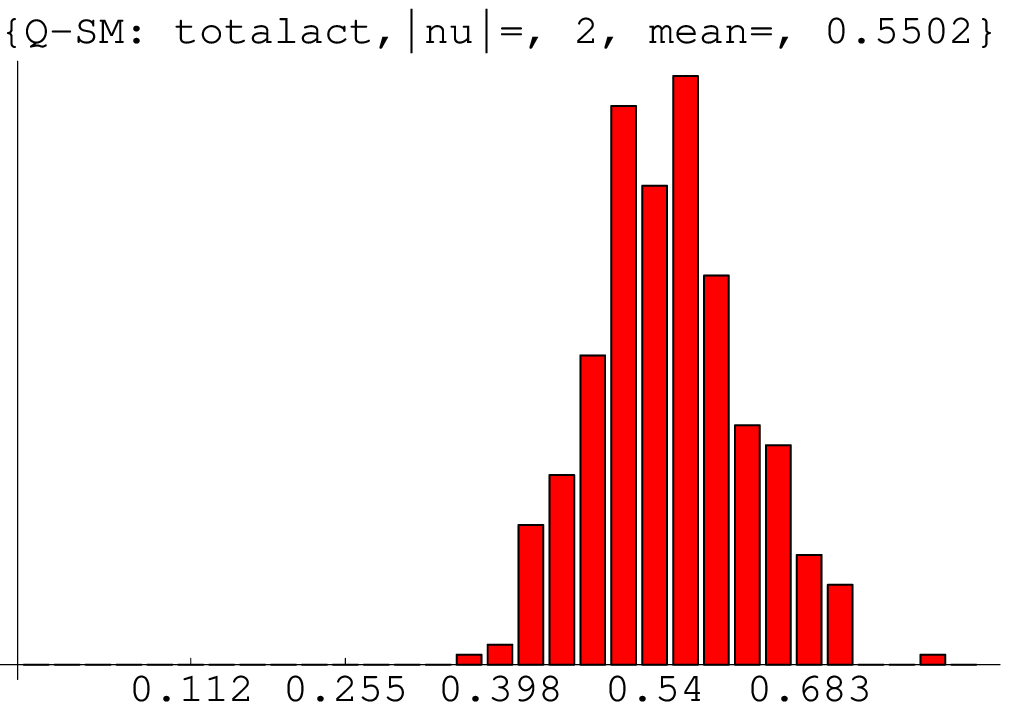,height=3.0cm,width=4.4cm,angle=0}
\vspace{-6mm}
\caption{\sl\small
Distribution of $S_{\rm gauge}$, $-\!\log(\det(D\!\!\!\!/+\!m))$ per continuum
flavour and $S_{\rm tot}$ [assuming two flavours] for $|\nu|\!=\!0,1,2$ in
the quenched theory ($\beta\!=\!4.0, m\!=\!0.05$).}
\end{figure}
In order to find out whether the `topological unquenching' shortcut has a
chance to provide a reasonable approximation to the full theory it seems
useful to check whether the underpinning assumption --a correlation between
$-\!\log(\det(D\!\!\!\!/+\!m))$ and $\int\!F\tilde F\,dx\,$-- holds true in the
full theory. The result which merely extends work initiated in the literature
\cite{Gattringeretal} is shown in Figs.~1 and~2.
From the scatter plots one notices that $\nu_{\rm nai}$ tends to accumulate
near integer values, both in the full and the quenched theories. The obvious
difference is that $|\nu|$ reaches up to~3 in the quenched approximation, but
up to~2 in the full theory, i.e.\ quenching results in a broadening of the
distribution of topological indices if $\beta$ is kept fixed.
The center of the clouds representing the gauge action in a given
topological sector shifts upwards, if $|\nu|$ increases. Regarding the
correlation between $-\!\log(\det(D\!\!\!\!/+\!m))$ and $\int\!F\tilde F\,dx$
one realizes that the center of the clouds at $\nu\!=\!\pm1$ in the central
plot on the l.h.s.\ of Fig.~1 is substantially higher than the center of the
cloud at $\nu\!=\!0$. It is interesting to note that the same correlation holds
true even in the quenched approximation, as one can see from the central plot
on the r.h.s.\ of Fig.~1. The difference is just that in the quenched updating
process this information hasn't been made use of.
In order to assess the shifts on a more quantitative level, the same
information is presented in Figs.~2 and~3 in the form of histograms for the
full and the quenched theories, respectively. They show the distribution of the
gauge action [per plaquette], of $-1/2$ times the logarithm of the staggered
determinant [per plaquette, after an overall shift to have it centered
at zero in the topologically trivial sector] and of the weighted sum [first
plus twice the second, $S_{\rm gauge}-2\log(\det(D\!\!\!\!/+\!m))=
S_{\rm gauge}-\log(\det(D_{\rm stag}))$]. The latter quantity represents the
total action the full sample was generated with, while in the quenched case it
is an artificial construct.
Considering the situation in the full theory first (Fig.~2), one realizes that
going from a configuration with $\nu\!=\!0$ to another one with
$\nu\!=\!\pm1,\pm2$ the gauge action per plaquette increases, on average, by
0.0134, 0.0213. In the same transition minus the log of the single-flavour
determinant per plaquette increases, on average, by 0.0082, 0.0157. Hence in
the simulation with a staggered pair the total action increases by 0.0299,
0.0527. This means that in the full theory the functional determinant is
more efficient in suppressing the higher topological sectors than the gauge
action.
Considering the analogous shifts in the quenched run (Fig 3.), a transition
$\nu\!=\!0\rightarrow\pm1,\pm2$ increases the gauge action per plaquette by
0.0038, 0.0309. Simultaneously the sign-flipped logarithm of the single-flavour
determinant per plaquette increases by 0.0032, 0.0100.
In spite of numerical values being somewhat different, the basic phenomenon
happens to be the same in either theory: Going from the lowest to the highest
sector, only the shift in the distribution of the log of the determinant is
comparable to its total width, the shift in the gauge action is almost
negligible on the scale of its total width.
Hence the correlation between $-\!\log(\det(D\!\!\!\!/+\!m))$ and $|\nu|$ seems
to be a robust phenomenon superseeding the choice of measure in the functional
integral.

\begin{figure}[h]
\epsfig{file=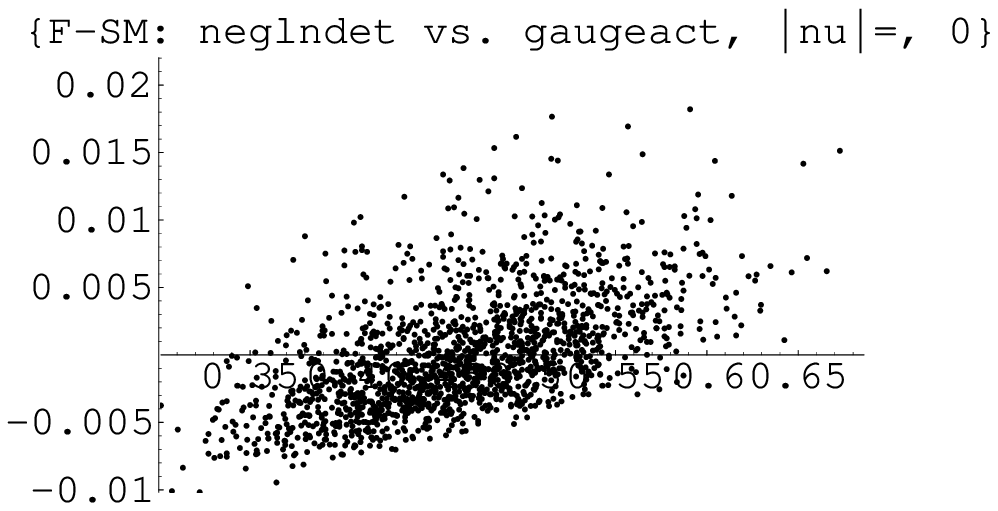,height=4.0cm,width=6.5cm,angle=0}
\hfill
\epsfig{file=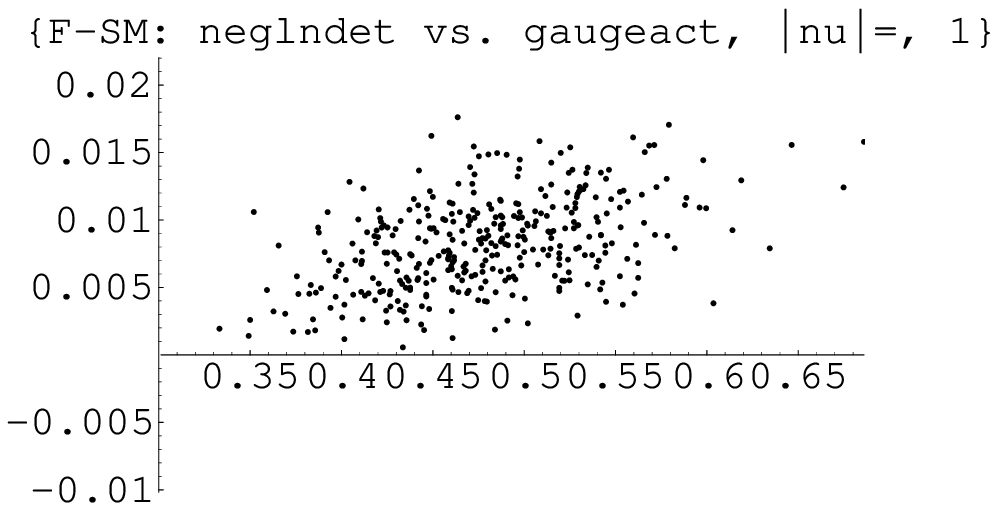,height=4.0cm,width=6.5cm,angle=0}
\vspace{2mm}
\\
\epsfig{file=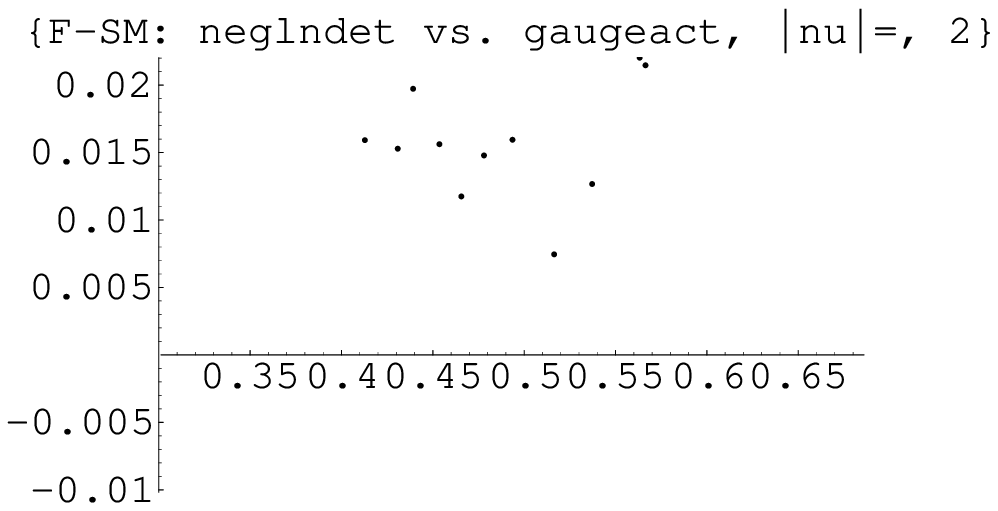,height=4.0cm,width=6.5cm,angle=0}
\hfill
\epsfig{file=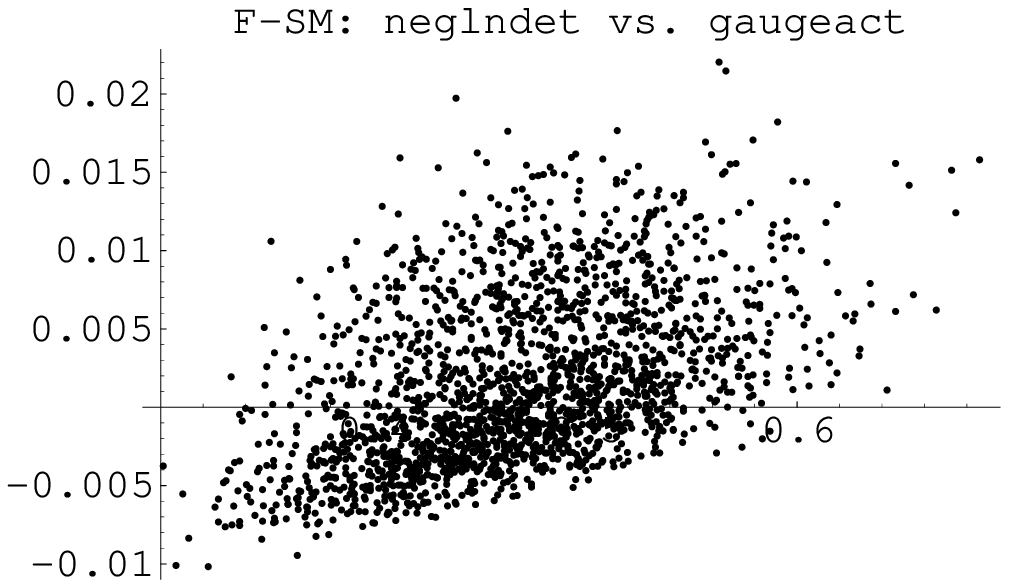,height=4.0cm,width=6.5cm,angle=0}
\vspace*{-4mm}
\caption{\sl\small
Scatter plots of $-\!\log(\det(D\!\!\!\!/\,+m))$ [per flavour] as a
function of $S_{\rm gauge}$ for $|\nu|\!=\!0,1,2$ and without separation into
topological sectors in the 2-flavour-theory.}
\end{figure}
From the discussion above it is clear that knowing the topological index
$\nu$ of a configuration allows for a (modestly accurate) `guess' of its
functional determinant. If, in addition, a correlation between
$-\!\log(\det(D\!\!\!\!/+\!m))$ and $\int\!F F\,dx$ holds true, a more accurate
guess would become possible. The result of an investigation in the full theory
on this topic is shown in Fig.~4. The first three plots show the sign-flipped
logarithm of the single-flavour determinant (normalized per plaquette) as a
function of the gauge action (per plaquette) for $\nu\!=\!0, \pm1, \pm2$.
One finds that --within a given topological sector-- these two quantities are
indeed correlated. Besides the vertical shift discussed above, a transition
$\nu\!=\!0\rightarrow\pm1$ implies a slight reduction in the slope of the
regression line. From this one concludes that, to a first approximation,
the functional determinant leads to a {\em reweighting of the topological
sectors} and to a {\em sectorally different renormalization of $\beta$}.
The last plot in Fig.~4 shows that, without separation  into different
topological sectors, the correlation is weaker and the statement that the
functional determinant results in an overall-renormalization of $\beta$ is
less useful.

\section{Implementing `topological unquenching'}

A practical issue in a `topologically unquenched' simulation is the strategy
adopted for building up the databases knowledge about
$\langle\,-\!\log(\det(D\!\!\!\!/+\!m))\,\rangle$ as function of $\nu$ in the
simple form of the method or as a function of $\nu$ and $S_{\rm gauge}$ in the
more elaborate form.
The point is that the desired correlation parameters [offset and slope in the
linear dependence of $\langle\,-\!\log(\det(D\!\!\!\!/+\!m))\,\rangle$ on
$S_{\rm gauge}$ for fixed $\nu$] are those in the full theory, but doing first
a run with the full measure is, of course, not an option.
The practical answer is that the `topologically unquenched' algorithm builds up
knowledge about the situation as it emerges in the `topologically unquenched'
run itself.
First one starts with a quenched pre-pre-thermalization, thereby acquiring
knowledge about offsets and slopes in the quenched theory.
\begin{figure}[h]
\epsfig{file=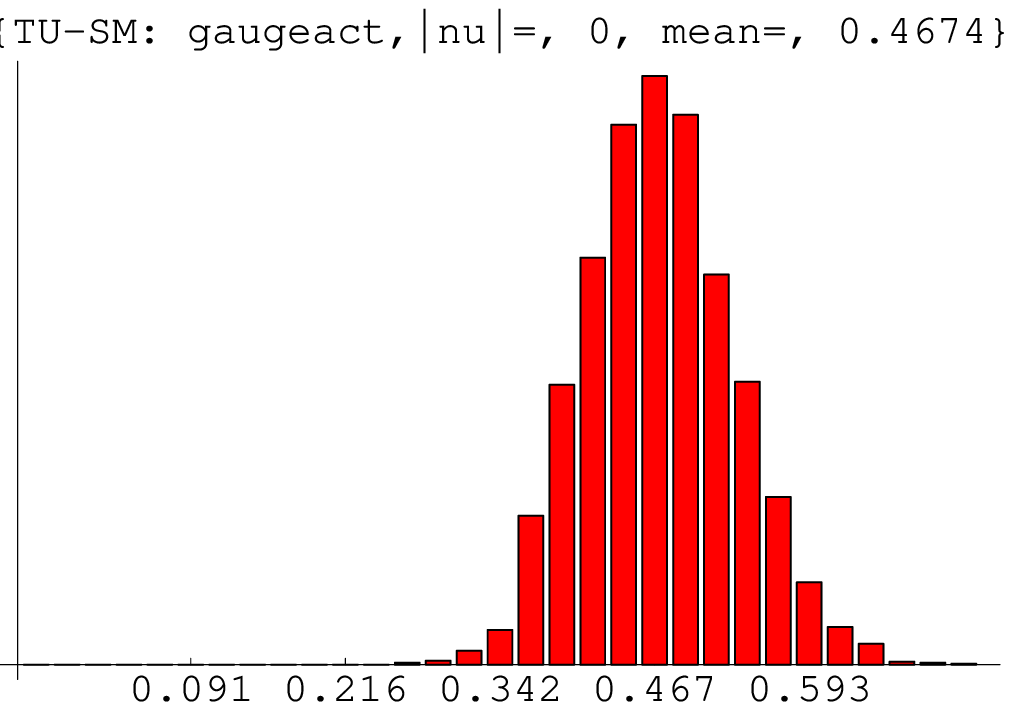,height=3.0cm,width=4.4cm,angle=0}
\hspace{0.5mm}
\epsfig{file=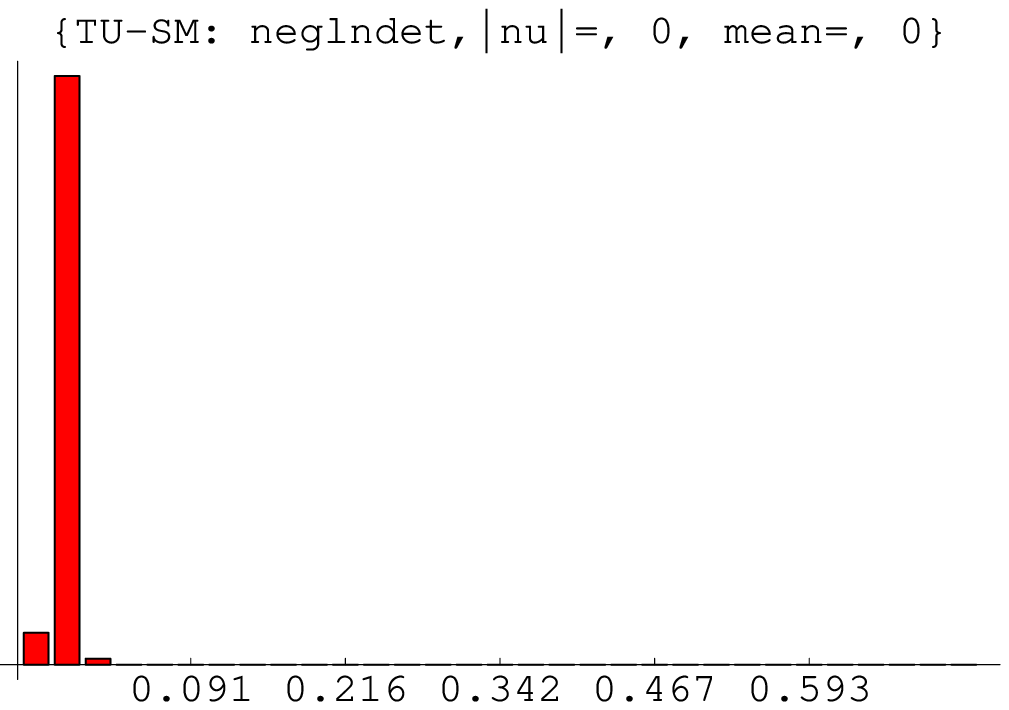,height=3.0cm,width=4.4cm,angle=0}
\hspace{0.5mm}
\epsfig{file=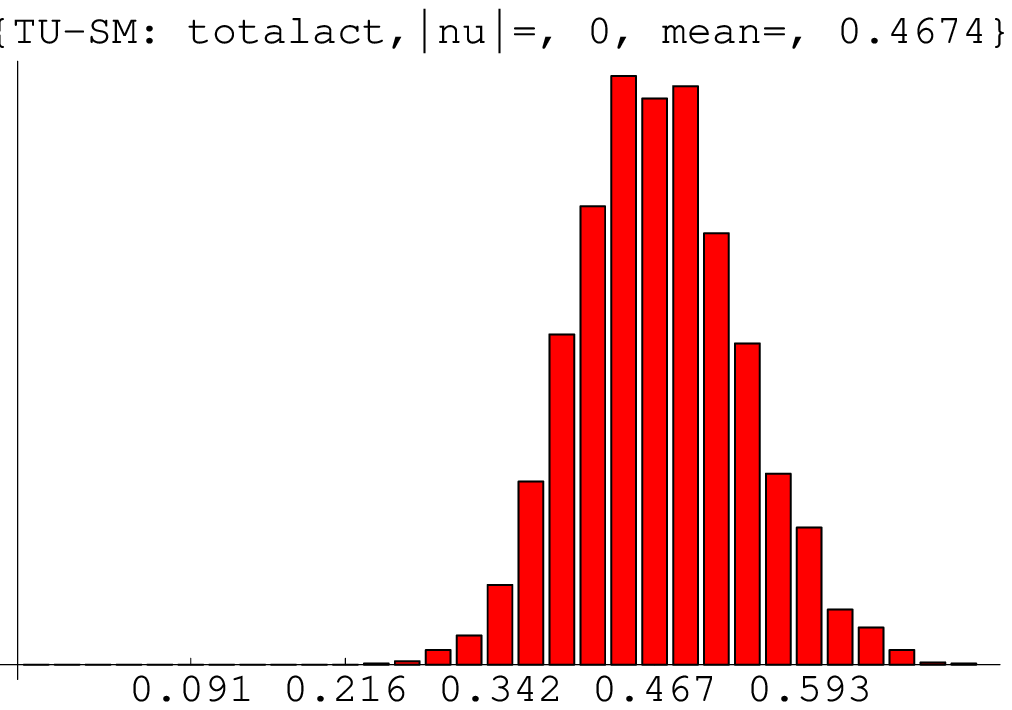,height=3.0cm,width=4.4cm,angle=0}
\vspace{-3mm}
\\
\epsfig{file=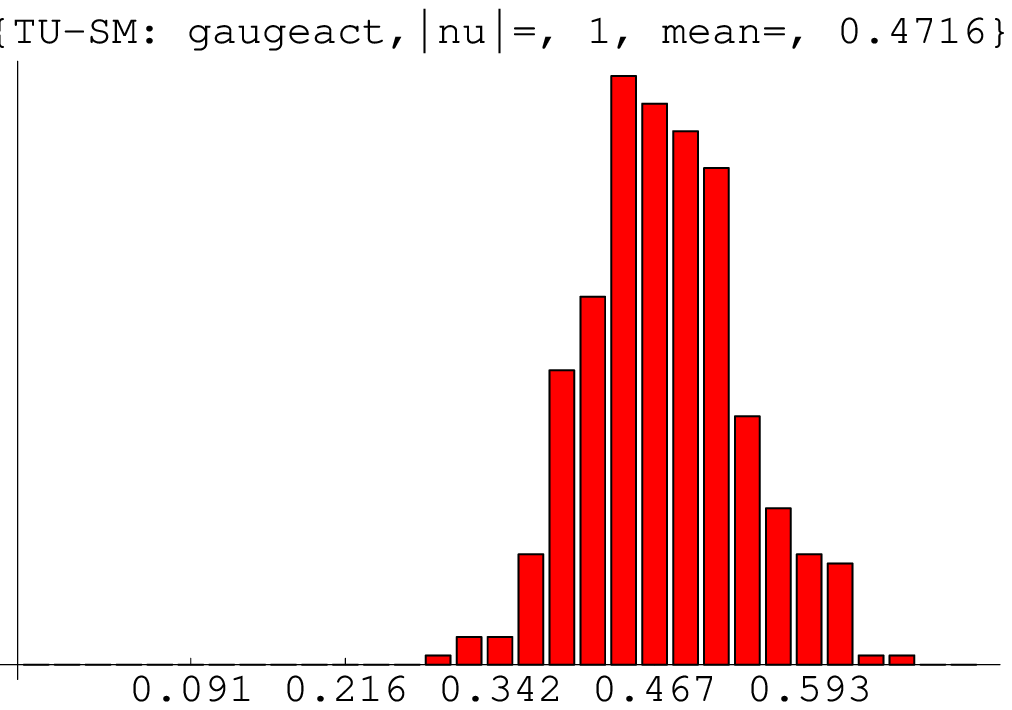,height=3.0cm,width=4.4cm,angle=0}
\hspace{0.5mm}
\epsfig{file=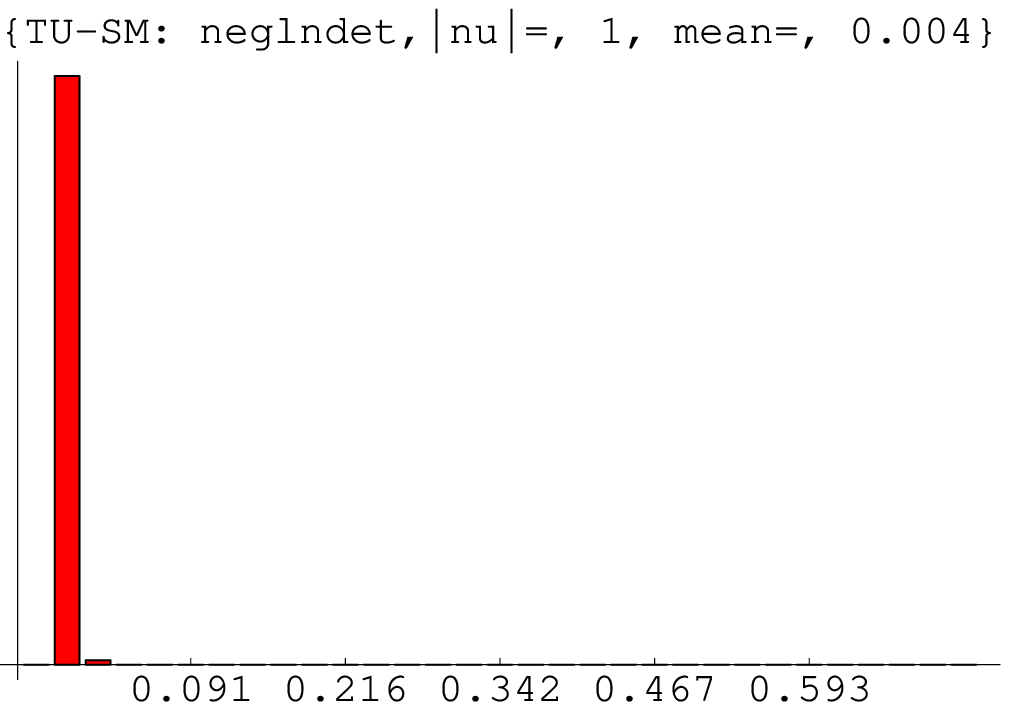,height=3.0cm,width=4.4cm,angle=0}
\hspace{0.5mm}
\epsfig{file=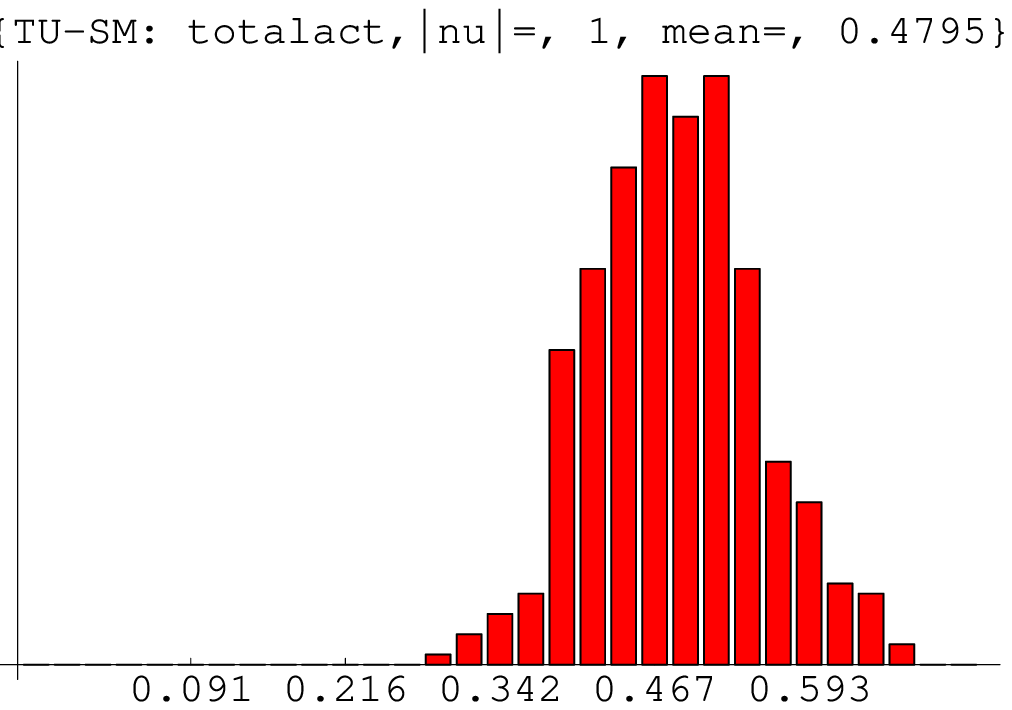,height=3.0cm,width=4.4cm,angle=0}
\vspace{-8mm}
\caption{\sl\small
Distribution of $S_{\rm gauge}$, $-\!\log(\det(D\!\!\!\!/+\!m))$ per continuum
flavour and $S_{\rm tot}$ for $|\nu|\!=\!0,1$ in the topologically unquenched
2-flavour theory ($\beta\!=\!4.0, m\!=\!0.05$).}
\end{figure}
Next follows a pre-thermalization phase where the number of `topologically
active' flavours increases, by small fractional steps, from zero to the
desired number of dynamical flavours. Accordingly, the databases values
for offsets and slopes will tend closer towards those appropriate in the
full theory.
Finally, the system is given time to thermalize with the true number of
`topologically unquenched' flavours, before actual measurements take place.
Fig.~5 shows the offsets for the centers of the distributions of the gauge
action, the sign-flipped logarithm of the single-flavour functional determinant
and the total action in a `topologically unquenched' 2-flavour system, data
being collected during the measurement phase.
As expected, numerical values lie in between those in the quenched and in the
full theory (see Figs.~2 and~3).

\section{Effect on selected observables}

In order to assess the implications for physics the `topological unquenching'
idea brings about, the static quark-antiquark potential has been determined
using the full, the quenched and the `topologically unquenched' measure.
\begin{figure}[h]
\epsfig{file=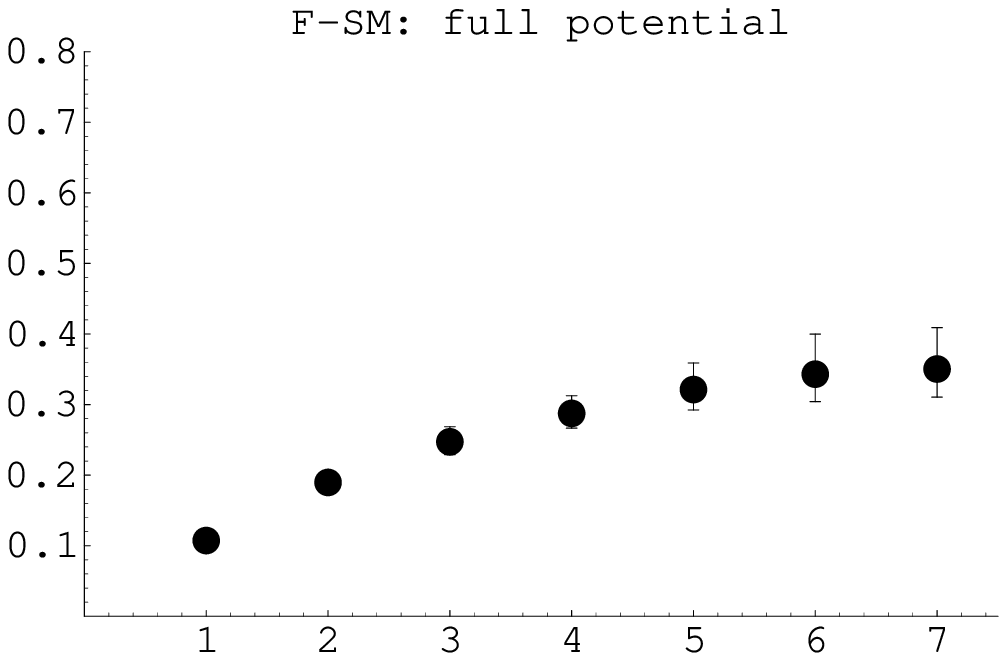,height=4.0cm,width=6.5cm,angle=0}
\hfill
\epsfig{file=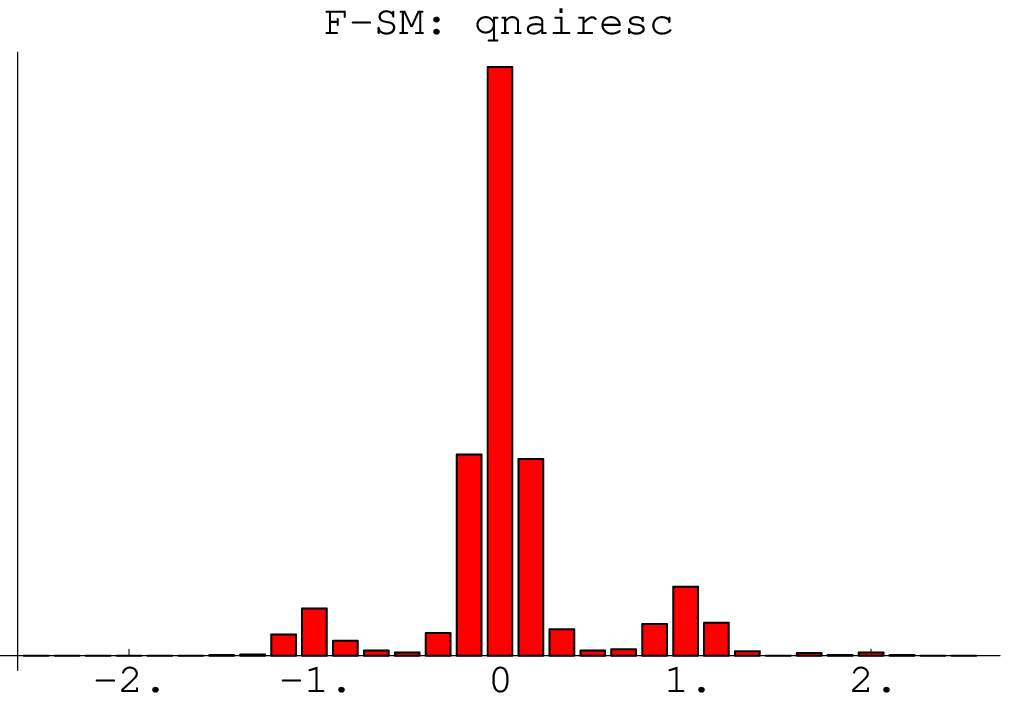,height=4.0cm,width=6.5cm,angle=0}
\vspace{2mm}
\\
\epsfig{file=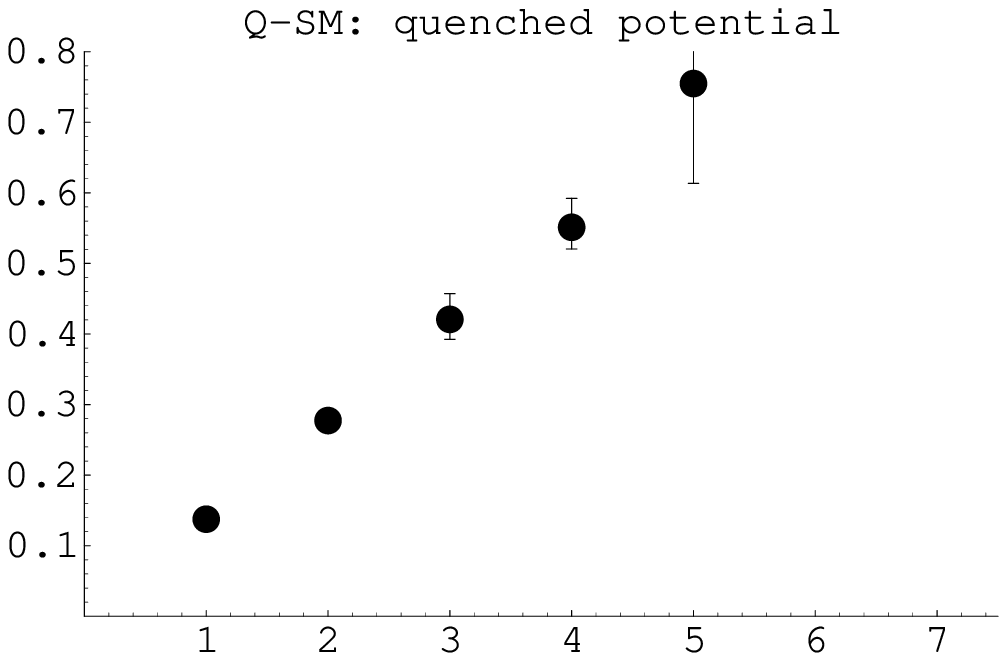,height=4.0cm,width=6.5cm,angle=0}
\hfill
\epsfig{file=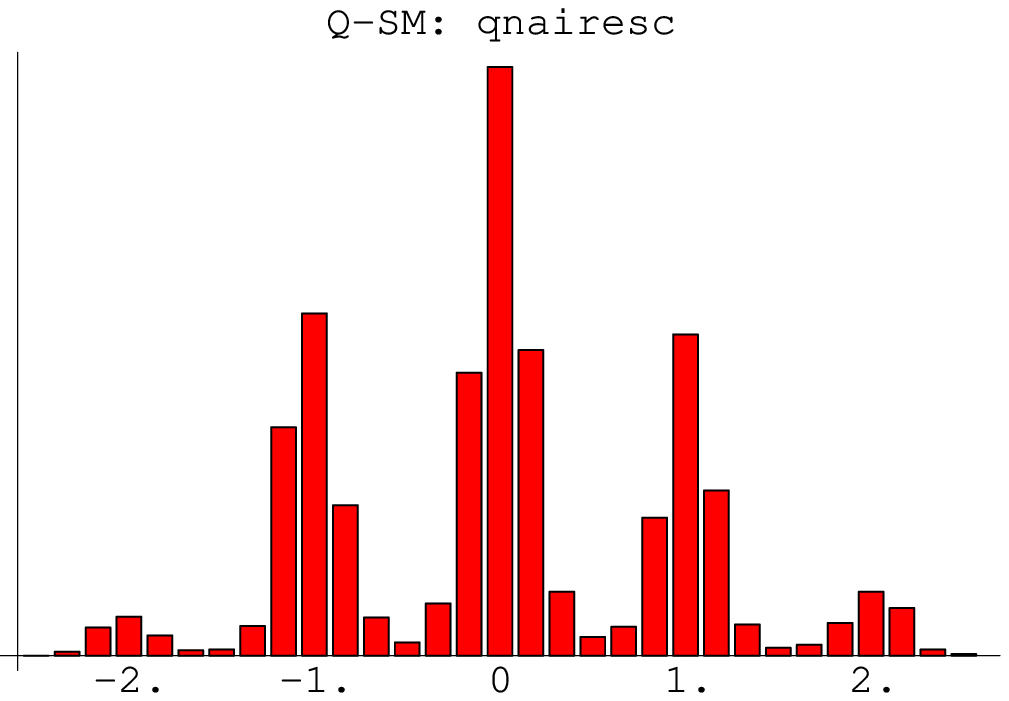,height=4.0cm,width=6.5cm,angle=0}
\vspace{2mm}
\\
\epsfig{file=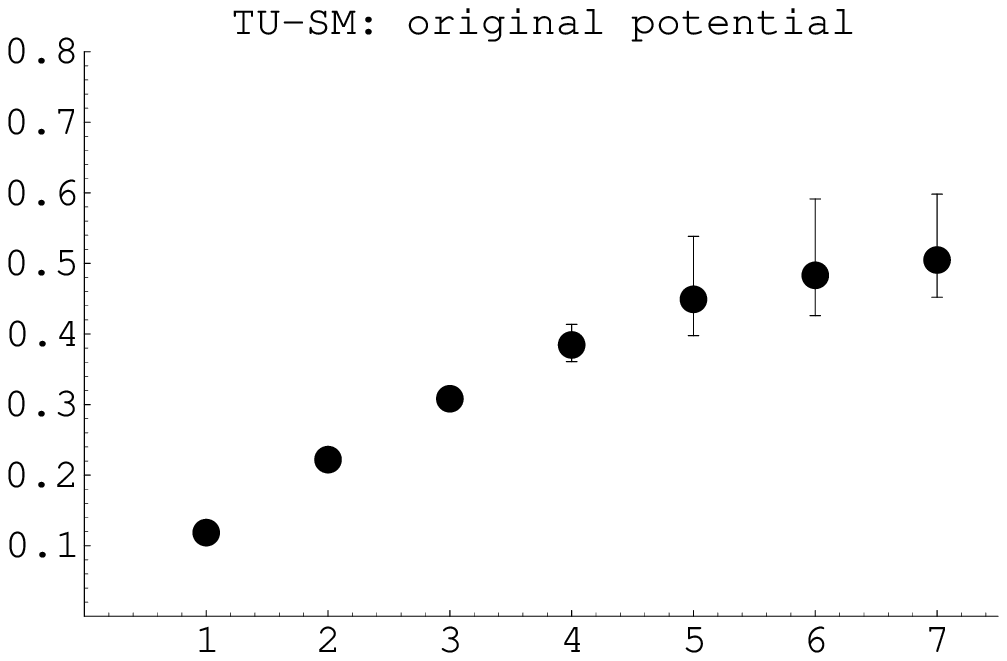,height=4.0cm,width=6.5cm,angle=0}
\hfill
\epsfig{file=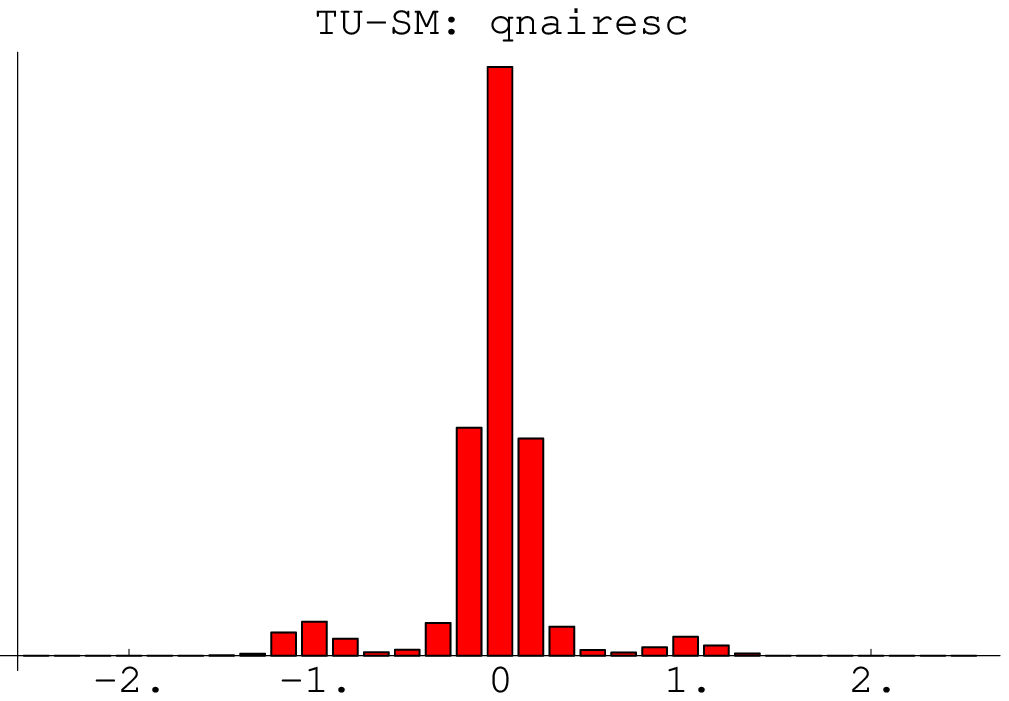,height=4.0cm,width=6.5cm,angle=0}
\vspace*{-4mm}
\caption{\sl\small
Static quark-antiquark potential and distribution of $\nu_{\rm nai}$ in the
full, quenched and `topologically unquenched' (without correction factor)
theories.}
\end{figure}
In the latter case, the more elaborate version has been used, i.e.\ the
database keeps track of the best linear fit for 
$\langle\,-\!\log(\det(D\!\!\!\!/+\!m))\,\rangle$ as 
a function of $\nu$ and $S_{\rm gauge}$.
The result is presented in the l.h.s.\ of Fig.~6.
As one can see, the quenched potential raises linearly\footnote{Unlike in QCD,
there is no $1/r$ short-distance tail.}, while in the full theory the test
charges get seriously screened at large distance.
The `topologically unquenched' recipe generates a sample in between; the
potential as it emerges from the unaltered sample shows insufficient yet
nontrivial screening at long distances.
What constitutes the primary difference between the samples is shown in
the r.h.s.\ of Fig.~6: The distribution of $\nu_{\rm nai}$ is much broader
in the quenched chase (middle) than in the full theory (top), while the
`topologically unquenched' run (bottom) nearly reproduces the full
distribution.

\begin{figure}
\epsfig{file=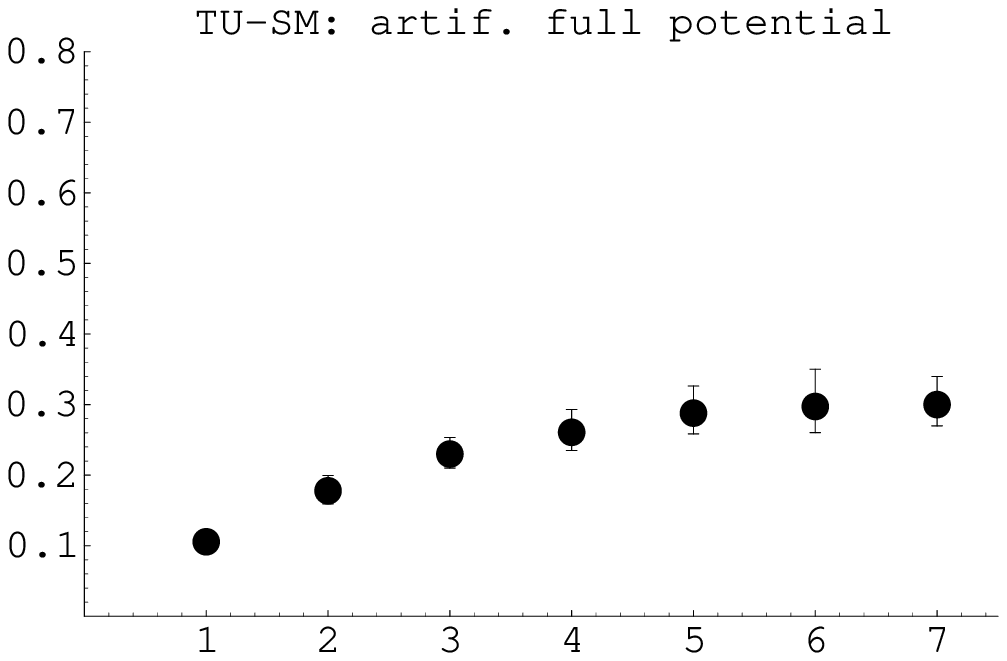,height=4.0cm,width=6.5cm,angle=0}
\hfill
\epsfig{file=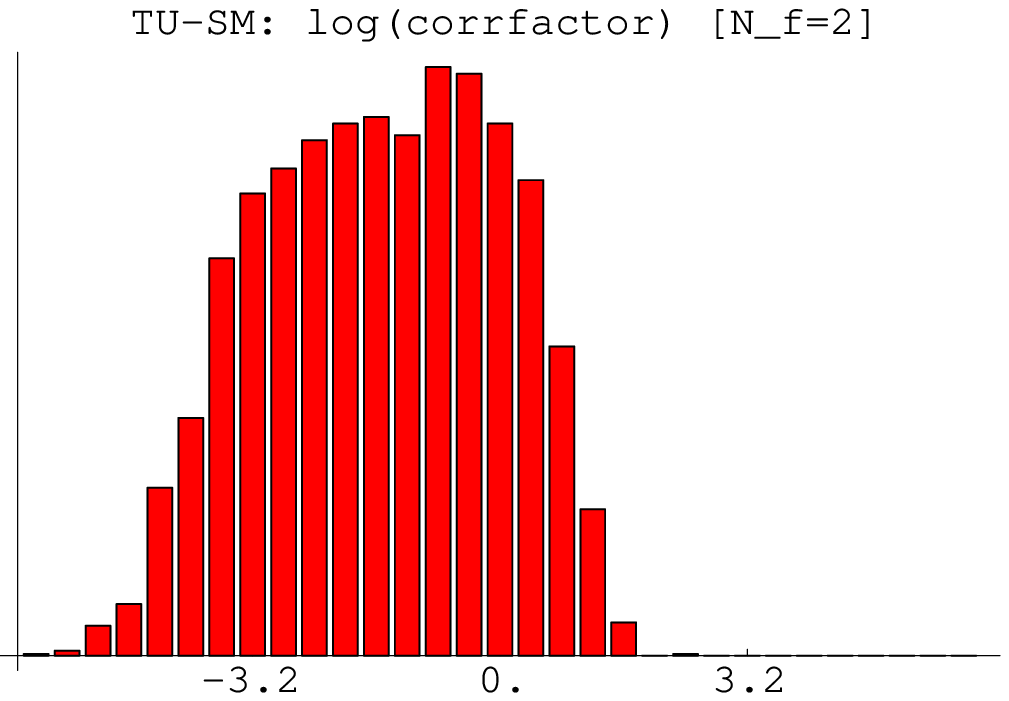,height=4.0cm,width=6.5cm,angle=0}
\vspace{-4mm}
\caption{\sl\small
Static quark-antiquark potential in the full theory as reconstructed from the
`topologically unquenched' run by including the correction factor into the
observable (left) and distribution of the $\log$ of the correction factor
(right).}
\end{figure}
Up to this point the option of eventually including the
`correction factor' [i.e.\ the $N_{\! f}$-fold power of the second factor in
(\ref{detfac})] into the observable has not been used.
An attempt to reconstruct the potential in the full theory from the
`topologically unquenched' sample by including this factor into the observable
yields the result shown in the l.h.s.\ of Fig.~7. A detailed comparison shows
that the potential agrees (within error bars) with the one obtained from the
full sample.
The only difference between the first plots in Fig.~6 and~7 is a technical one:
The full potential reconstructed from the `topologically unquenched' sample
was cheaper, in terms of CPU time, by a factor $\sim250$ than the one won from
the full sample.
It should be mentioned, however, that it was not clear a priori that including
the correction factor into the observable would be an option.
The point is that the correction factor is the ratio of two determinants
each of which scales exponentially with the total lattice volume.
This means that the extra fluctuations brought by the `correction factor'
grow exponentially as the lattice gets large.
In the present example the distribution of the log of the correction factor
is sufficiently narrow as to allow for a reconstruction of the full potential.
As one can see from the r.h.s.\ of Fig.~7 the majority of the configurations
has a log between -3.2 and 0.0, i.e.\ most configurations receive a reweighting
factor between $\sim\!0.1$ and 1.
At fixed $\beta, m$ this width grows linearly with $V$, thus making such
a reweighting impossible for sufficiently large $V$.
It should be emphasized, however, that even on a large lattice the part of
the functional determinant which has been absorbed into the `topologically
unquenched' measure results --for most observables-- in an improvement on
the quenched approximation (see Fig.~6 again).

An extreme `counterexample' where the part of the functional determinant
which has been absorbed into the `topologically unquenched' measure does
not result in any improvement at all is shown in Fig.~8.
The expectation value of the Polyakov line is still zero, like in the quenched
theory, whereas in the full theory it is real and positive.
In this case the entire shift between the quenched and the full theories
emerges from the factor which, in the `topological unquenching' framework, is
considered a `correction factor'. It should be noted, however, that this
particular test is extremely hard to pass, because the difference between
the quenched and the full theories is much more severe than in the case of QCD.
This holds true because the quenched Schwinger model is always in the confined
phase, while the full multiflavour Schwinger model is always in the deconfined
phase -- regardless of the actual temperature. In other words: The effect of
the determinant in the massive multiflavour Schwinger model is not merely
renormalizing masses and lengths; quenching results in qualitative changes for
any values of $\beta, m$, making it a more drastic approximation than in QCD.
\begin{figure}
\epsfig{file=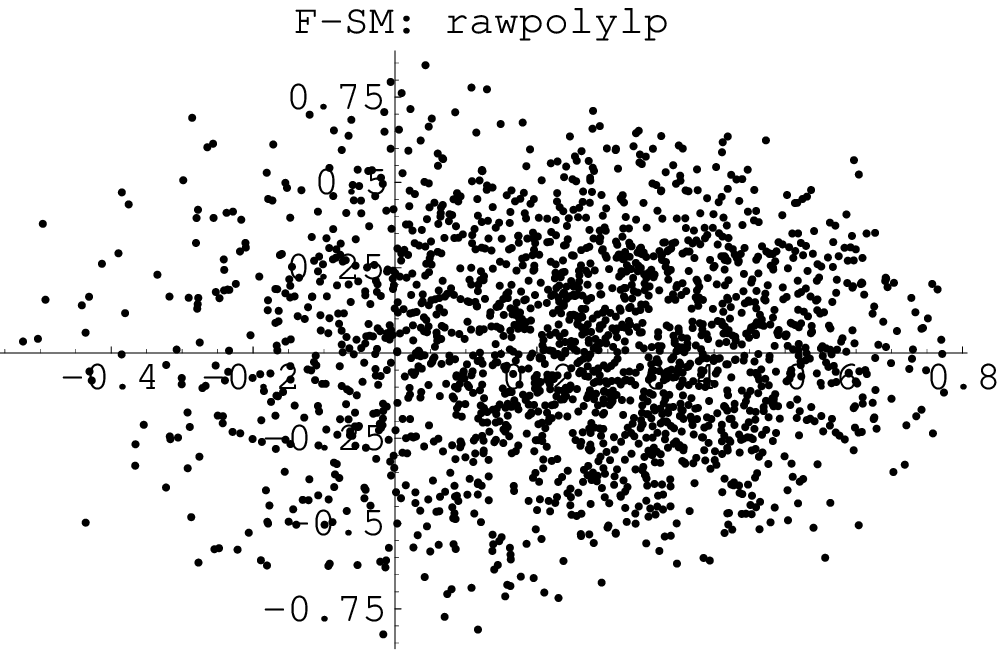,height=4.0cm,width=6.5cm,angle=0}
\hfill
\epsfig{file=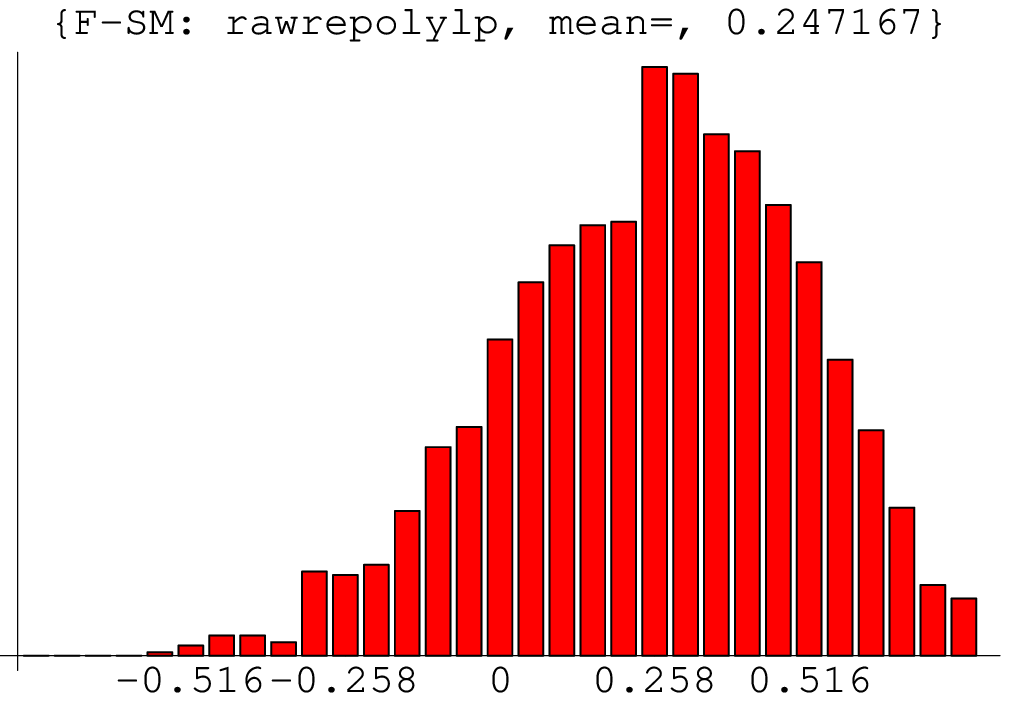,height=4.0cm,width=6.5cm,angle=0}
\vspace{2mm}
\\
\epsfig{file=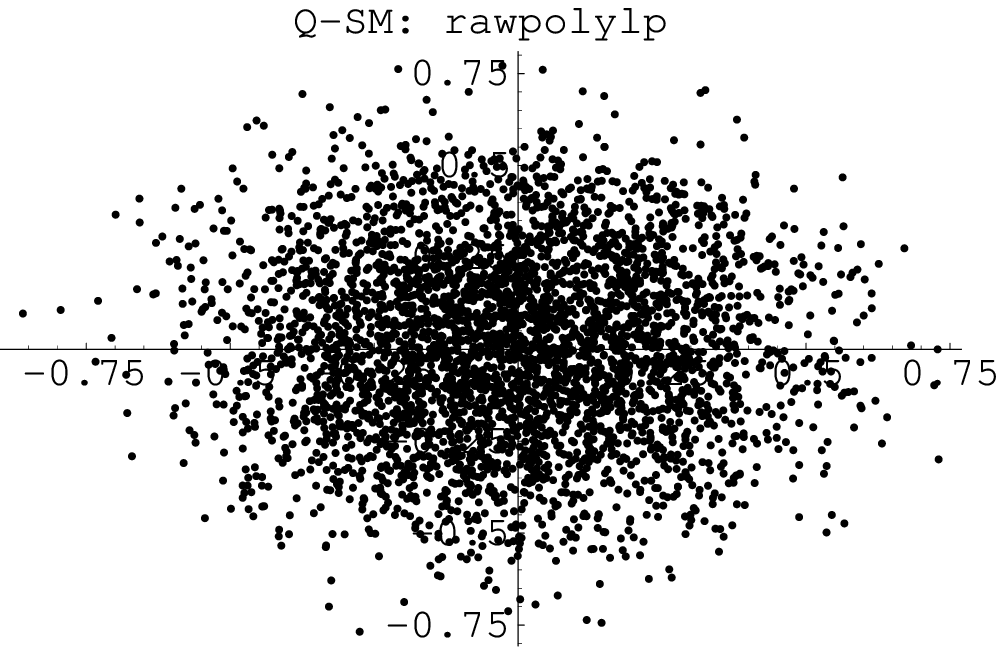,height=4.0cm,width=6.5cm,angle=0}
\hfill
\epsfig{file=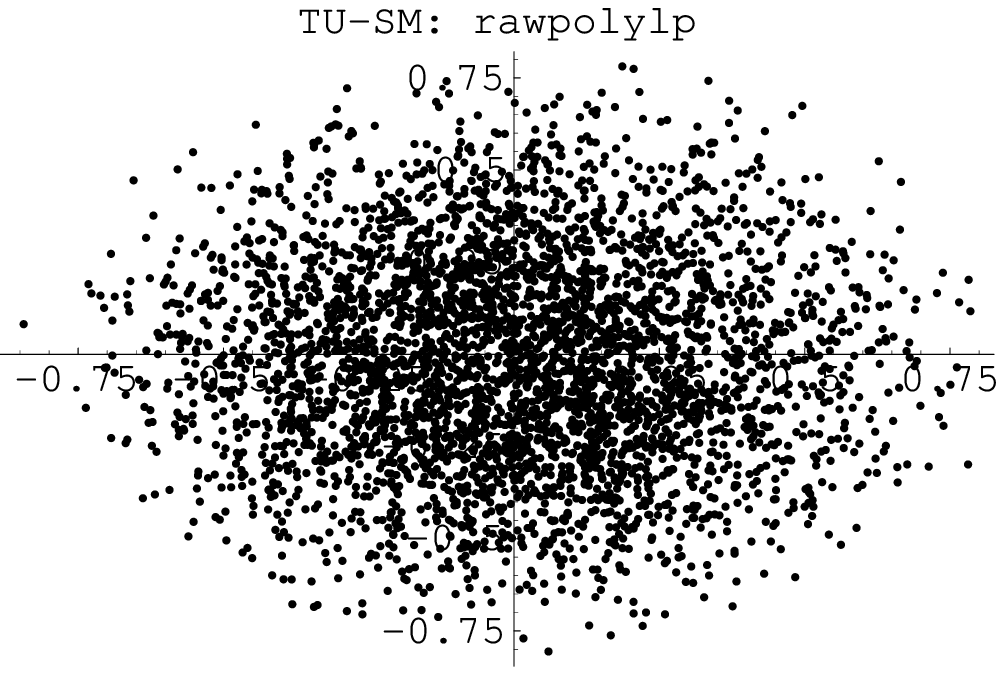,height=4.0cm,width=6.5cm,angle=0}
\vspace{-2mm}
\caption{\sl\small
Scatter plot of the Polyakov line and distribution of its real part
in the full theory (top); Polyakov line in the quenched and the topologically
unquenched (without correction factor) theories (bottom left and right).}
\end{figure}

It is interesting to elucidate the physical reason why --at least for the
standard observable of the heavy quark potential-- `topological unquenching'
results in a substantial improvement of the behaviour at long distance.
To this end the configurations generated in the full theory have been
re-analyzed in different classes, according to the absolute value of their
topological index $\nu$. As one can see from Fig.~9, the `sectoral' potentials
for $|\nu|\!=\!0$ and $|\nu|\!=\!1$ differ quite drastically.
Comparing the l.h.s.\ of Fig.~9 to the first plot in Fig.~6 one notices
that the sectoral potential with $|\nu|\!=\!0$ almost agrees with the
unseparated `true' potential. The big surprise, however, is that the error bars
are smaller now, even though the `topologically trivial' potential emerges from
a subset of the configurations used for the full potential. It seems that in
this particular case the higher sectors primarily add some `noise'. It should
be noted, however, that the topologically trivial sector does not always yield
the right result. In a previous run in a (physically) larger box, the
`topologically trivial' potential was found to lie substantially below the true
potential \cite{LAT99}, i.e.\ a strict constraint to the trivial sector would
not always prove beneficial.
This is reminiscent of the situation in QCD with $N_{\!f}\!\geq\!2$, where
Leutwyler and Smilga have shown that in a sufficiently small box
($x\!=\!V\Sigma m/N_{\! f}\!\ll\!1$) the partition function is dominated by
the topologically trivial sector, whereas in a sufficiently large box
($x\!\gg\!1$) a global quantity like $\int\!F\tilde F\,dx$ has vanishing
influence on physical variables \cite{LeutSmil}.
In this language the current simulation is in the regime $x\!\ll\!1$ whereas
the previous one \cite{LAT99} should likely be associated\footnote{In the
context of the multiflavour Schwinger model $\Sigma (\!=\!0)$ should be traded
for $F_\pi (\!\neq\!0)$.} with $x\!\simeq\!1$.
From the combined simulation data it is natural to conjecture that, if the
box-volume increases gradually so as to leave the regime $x\!\ll\!1$,
{\em several mutually inconsistent sectors\/} start to yield {\em sizable
contributions\/}; only as $x$ goes beyond~1, the differences between the
sectoral contributions start fading away.
Since the regime $x\simeq\!1$ is not uncommon in current lattice simulations,
it seems desirable to respect the relative weight of the different topological
sectors in the full theory -- even if importance sampling is done w.r.t.\ to a
different measure.
\begin{figure}
\epsfig{file=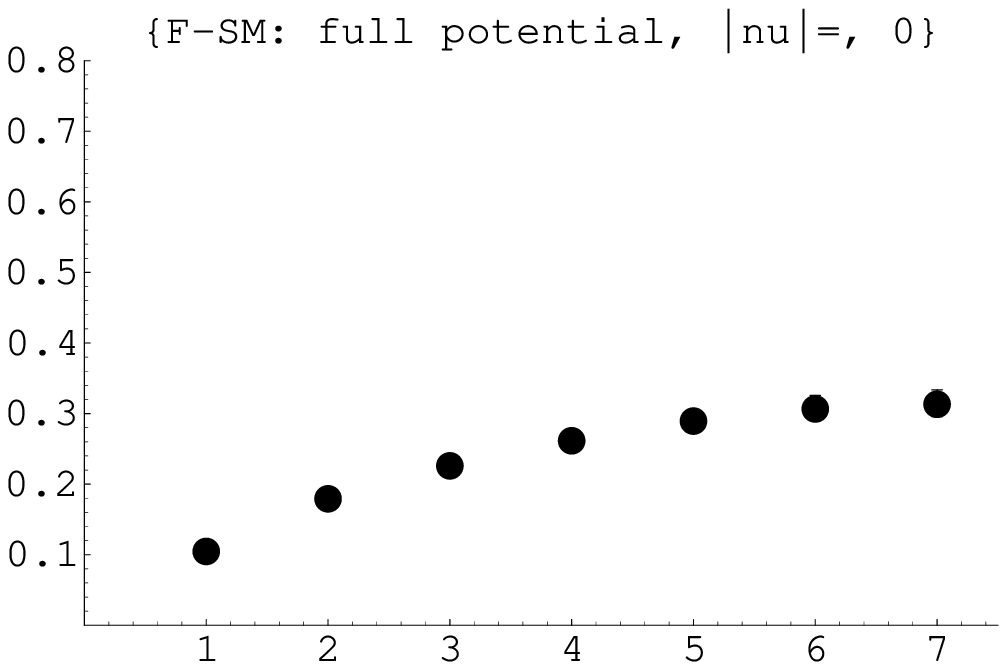,height=4.0cm,width=6.5cm,angle=0}
\hfill
\epsfig{file=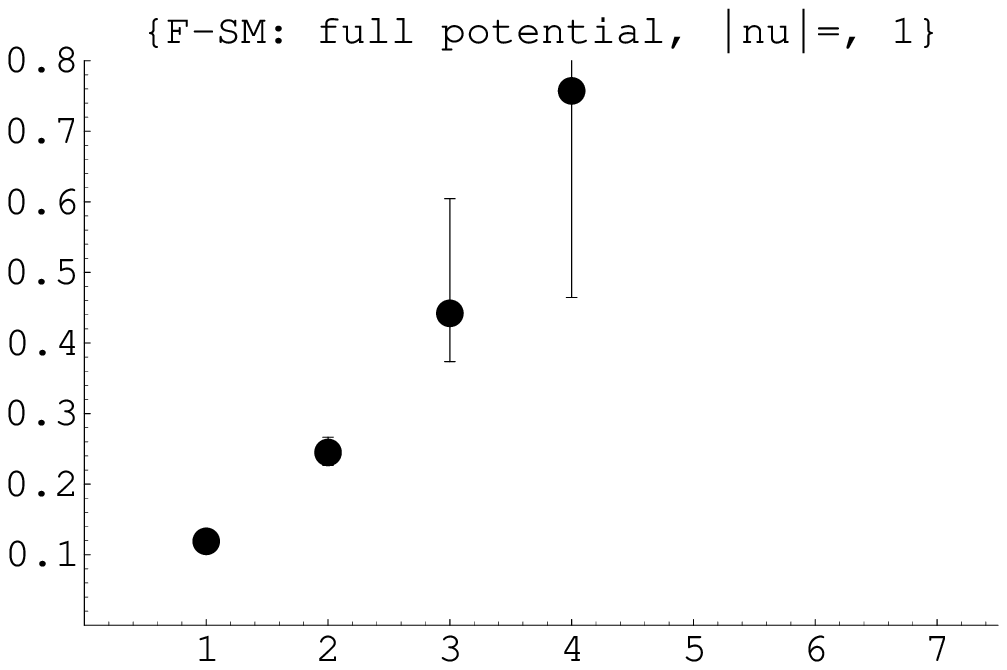,height=4.0cm,width=6.5cm,angle=0}
\vspace{-4mm}
\caption{\sl\small
Potential in the full theory, if only configurations with $\nu\!=\!0$ (left)
or $\nu\!=\!\pm1$ (right) are evaluated.}
\end{figure}

Summarizing our investigations in the massive multiflavour Schwinger model, one
might say that the basic concept of `topological unquenching' --including the
first factor in (\ref{detfac}) into the measure-- results for several
observables in a substantial improvement on the quenched approximation at
virtually no costs, whereas the option of including the second factor into
the observable is likely to be limited to very small lattices.

\section{Extension to QCD}

Finally, a few comments and speculations on a possible extension of the
`topological unquenching' idea to QCD are in order.

It is clear that the basic feature the approach relies on --a correlation
between the logarithm of the determinant and one or several global gluonic
quantities which may be kept track of from looking at their respective local
changes-- holds true in the case of QCD too.
What is not so clear is whether this correlation is strong enough as to
make this a practical way of guessing the change in the logarithm of the
determinant if a link in a phenomenological simulation is updated.
Of course, there is the possibility that further quantities, besides
$\int\!F\tilde F\,dx$ and $\int\!FF\,dx$, are found which correlate with
the determinants logarithm, thus allowing for a more accurate guess.

An obvious question, however, is which definition for the $F\tilde F$ operator
should be chosen on the lattice, since, with any practical implementation,
$\int\!F\tilde F\,dx$ is not an integer.
To make the situation even worse established techniques to render a
configuration more continuum-like [e.g.\ blocking inverse-blocking cycles,
link fuzzing or cooling] are known to cause any field-theoretic definition of
the topological charge `jump' one or several times about roughly integer values
until it finally settles close to an integer.
From a physical point of view, however, such a stability requirement is not
really useful as it leads to a determination of the topological charge in a
configuration which agrees in its instanton content with the original one only
modulo instantons smaller than a certain critical radius $\rho_{\rm crit}$, as
discussed in the appendix of \cite{TUQCD}.
It is thus preferable to apply as little cooling as possible and to allow for
some uncertainty, but to strive for the topological charge of the original
configuration. In other words: In the `topological unquenching' context there
is no need for the peaks and valleys in the $\nu_{\rm nai}$-distribution to be
so much pronounced as in the beautyful plot shown in Fig.~10. Any determination
which results in a distribution which takes values in the valleys not more than
half of the values in the neighboring peaks will do fine for `topological
unquenching' purposes. If improved operators are used, both $FF$ and
$F\tilde F$ should be dealt with in comparable manner. It is believed that even
with moderate improvement one or two cooling sweeps will be sufficient to
satisfy the valley-peak-criterion for current $\beta$-values. In practical
terms more cooling sweeps would be unpleasant anyways, since 2 cooling sweeps
require permanently carrying along the 1- and 2-cooled versions of the actual
configuration in order to minimize time spent on local cooling, when an
individual link is updated. It should be noted, however, that the situation
improves with increasing $\beta$. Eventually, for sufficiently high $\beta$,
even the Wilson action and the corresponding unimproved $F\tilde F$ operator
will satisfy the criterion, thus allowing for a way to keep track of
$\int\!F\tilde F\,dx$ which does not require cooling at all.
\begin{figure}
\begin{center}
\vspace{-14mm}
\includegraphics{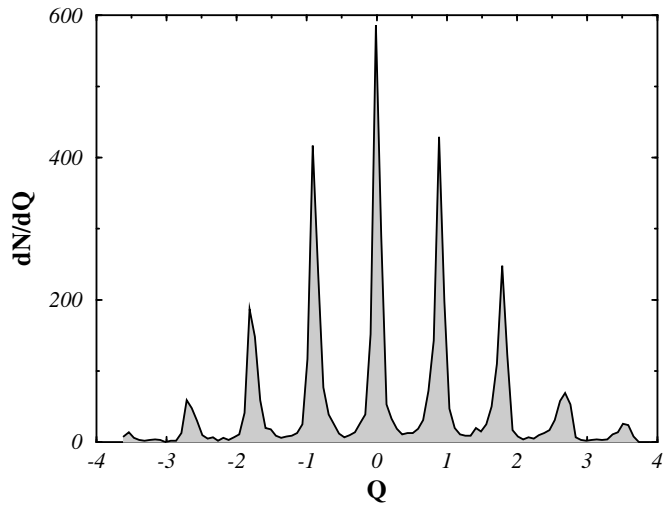}
\end{center}
\vspace{72mm}
\caption{\sl\small
Distribution of topological charge $\nu_{\rm nai}/\kappa_{\rm nai}$ (i.e.\
prior to rescaling) for an ensemble of 5000 $SU(3)$-configurations generated
from $S_{\rm Wilson}$ at $\beta\!=\!6.1$, after 6 cooling sweeps and with a
2-smeared charge operator. Figure taken from \cite{Alles}.}
\end{figure}

Regardless of how much local cooling is needed for $\beta$-values used
over the next few years, what is important is the following:
Since the computational overhead of the `topologically unquenched' over the
quenched approximation stems entirely from local efforts to keep track of the
topological charge, costs per configuration get increased by a fixed
factor\footnote{It depends on the volume of the neighborhood local cooling is
applied to, which in turn depends on the locality properties of
$FF(x), F\tilde F(x)$ and the number of cooling cycles. With 2 sweeps and not
more than next-to-nearest neighbor couplings, it might be $O(30)$.}
which is independent of the size of the lattice.
It is also instructive to consider the effect of accidental
`misidentifications' of the topological charge on the `topologically
unquenched' sample: In the extreme case where the charge determination yields a
normal distributed\footnote{This is what the distribution shown in Fig.~10
tends to, if the peaks and valleys get washed out completely. For an account on
the overall-distribution see Ref.~\cite{LeutSmil}.} random number which has
nothing to do with the actual configuration the simple version of the algorithm
finds no correlation between $-\!\log(\det(D\!\!\!\!/+\!m))$ and
$\int\!F\tilde F\,dx$ and produces, for this reason, a quenched sample.
Under the same circumstances the more elaborate version finds just a modest
linear correlation between $-\!\log(\det(D\!\!\!\!/+\!m))$ and $\int\!FF\,dx$
and produces, for this reason, a quenched sample with an
overall-rescaled $\beta$ (see Fig.~4, last plot).
\enlargethispage{8pt}

\section{Summary}

In this paper the idea of `topological unquenching' has been transformed into
a general scheme applicable to QCD inspired theories: The algorithm tries
to `guess' the change in the determinant a proposed link update would bring.
This is done using statistical information how the logarithm of the determinant
correlates with global gluonic quantities like $\int\!F\tilde F\, dx$ and
$\int\!FF\,dx$ which may easily be kept track of by looking at their respective
local changes.
Compared to the quenched approximation `topological unquenching' seems to bring
considerable improvement if the observable under consideration depends
sensitively on the structure of the long-range Green's functions in the full
theory and if, in addition, the simulation is done in the small
($V\Sigma m/N_{\!f}\!\ll\!1$) or intermediate ($V\Sigma m/N_{\!f}\!\simeq\!1$)
Leutwyler Smilga regime.

Obviously, a test-implementation like the one presented in this study cannot
prove that the idea of `topological unquenching' would be successful in the
case of QCD.
Nevertheless, the method seems interesting because of its sampling speed
and its potential field-theoretic virtues: Even an expensive formulation
respecting chiral symmetry may be used for the sea-quarks and the latter do
naturally have exactly the same mass as the current-quarks.
It is hoped that these features look promising enough as to persuade at least
one lattice group to give `topological unquenching' a try for the case of QCD.

\bigskip\noindent{\bf Acknowledgements}:
Long-term discussions with Stephen R. Sharpe over the issue presented in this
work are gratefully acknowledged.

\end{document}